\documentclass[10pt,twocolumn,twoside]{IEEEtran}
\renewcommand{\baselinestretch}{0.92}

\usepackage{tcolorbox}
\usepackage{graphicx}
\usepackage{xcolor}
\usepackage{theorem}
\usepackage{times,amsmath,epsfig}
\usepackage{amssymb}
\usepackage{subfigure}
\usepackage{cite}
\usepackage{cases}
\usepackage{url}

\usepackage{algorithmic}
\usepackage{algorithm}
\input{mysymbol.sty}

\newcommand{\EE}{{\mathbb E}}

\newtheorem{mycorollary}{\bf Corollary}
\newtheorem{mylemma}{\bf Lemma}
\newtheorem{myproposition}{\bf Proposition}
\newtheorem{remark}{\bf Remark}

\newtcolorbox{myblockt}[1]{colback=urblue!5!white,
	colframe=urblue,fonttitle=\bfseries,
	title=#1}

\newtcolorbox{myblock}{colback=urblue!5!white,
	colframe=urblue,fonttitle=\bfseries}

\title{Identifying the Topology of Undirected Networks from Diffused Non-stationary Graph Signals}

\author{\IEEEauthorblockN{Rasoul Shafipour, \emph{Student Member, IEEE}, Santiago Segarra, \emph{Member, IEEE,}\\  Antonio G. Marques, \emph{Senior Member, IEEE,} and Gonzalo Mateos, \emph{Senior Member, IEEE}}
\thanks{Work in this paper is supported by the NSF award CCF-1750428, the Spanish MINECO grants OMICRON (TEC2013-41604-R) and KLINILYCS (TEC2016-75361-R), and MIT IDSS seed grant. R. Shafipour and G. Mateos are with the Dept. of Electrical and Computer Eng., Univ. of Rochester. S. Segarra is with the Dept. of Electrical and Computer Eng., Rice University.  A. G. Marques is with the Dept. of Signal Theory and Comms., King Juan Carlos Univ.  Emails: rshafipo@ur.rochester.edu, segarra@rice.edu, antonio.garcia.marques@urjc.es, and gmateosb@ece.rochester.edu. Part of the results in this paper were presented at the \textit{2017 and 2018 ICASSP Conferences}~\cite{RSSSAMGM_icassp17,RSSSAMGM_icassp18}.}}

\begin{document}
\maketitle

\begin{abstract}%
We address the problem of \textit{inferring} an undirected \textit{graph} from nodal observations, which are modeled as \emph{non-stationary} graph signals generated by local diffusion dynamics that depend on the structure of the unknown network. Using the so-called graph-shift operator (GSO), which is a matrix representation of the graph, we first identify the eigenvectors of the shift matrix from realizations of the diffused signals, and then estimate the eigenvalues by imposing desirable properties on the graph to be recovered. Different from the stationary setting where the eigenvectors can be obtained directly from the covariance matrix of the observations, here we need to estimate first the unknown \textit{diffusion} (graph) \textit{filter} -- a polynomial in the GSO that preserves the sought eigenbasis. To carry out this initial system identification step, we exploit different sources of information on the arbitrarily-correlated input signal driving the diffusion on the graph. We first explore the simpler case where the observations, the input information, and the unknown graph filter are linearly related. We then address the case where the relation is given by a system of matrix quadratic equations, which arises in pragmatic scenarios where only the second-order statistics of the inputs are available. While such quadratic filter identification problem boils down to a non-convex fourth-order polynomial minimization, we discuss identifiability conditions, propose algorithms to approximate the solution and analyze their performance. {Numerical tests illustrate the effectiveness of the proposed topology inference algorithms in recovering brain, social, financial and urban transportation networks using synthetic and real-world signals.} 
\end{abstract}


%
\section{Introduction}\label{S:Introduction}

Consider a network represented as a weighted and undirected graph $\ccalG$, consisting of a node set $\ccalN$ of cardinality $N$, an edge set $\ccalE$ of unordered pairs of elements in $\ccalN$, and edge weights $A_{ij}\in\reals$ such that $A_{ij}=A_{ji}\neq 0$ for all $(i,j)\in\ccalE$. The edge weights $A_{ij}$ are collected in the \textit{symmetric} adjacency matrix $\bbA\in\reals^{N\times N}$. More broadly, one can define a generic \emph{graph-shift operator} (GSO) $\bbS\in\reals^{N\times N}$ as any matrix having the same sparsity pattern than that of $\ccalG$~\cite{SandryMouraSPG_TSP13}. Although the choice of $\bbS$ can be adapted to the problem at hand, it is often chosen as either $\bbA$, the Laplacian $\bbL:=\diag(\bbA\bbone)-\bbA$, or its normalized counterparts~\cite{EmergingFieldGSP}. 

Our focus in this paper is on identifying graphs that explain the structure of a random signal.
Formally, let $\bbx=[x_1,...,x_N]^T \in\mbR^N$ be a zero-mean graph signal with covariance matrix $\bbC_\bbx=\E{\bbx\bbx^T}$, in which the $i$th element $x_i$ denotes the signal value at node $i$ of an \emph{unknown graph} $\ccalG$ with shift operator $\bbS$. We say that the graph $\bbS$ represents the structure of the signal $\bby\in\mbR^N$ if there exists a diffusion process in the GSO $\bbS$ that produces the signal $\bby$ from the input signal $\bbx$, that is
\begin{equation}\label{eqn_diffusion}
	 \bby  = \textstyle  \alpha_0 \prod_{l=1}^{\infty} (\bbI_N-\alpha_l \bbS) \bbx
	  = \sum_{l=0}^{\infty}\beta_l \bbS^l\,\bbx,
\end{equation}
for some set of parameters $\{\alpha_l\}$ or, equivalently, $\{\beta_l\}$.
While $\bbS$ encodes local one-hop interactions, each successive application of the shift in \eqref{eqn_diffusion} percolates $\bbx$ over $\ccalG$; see e.g.~\cite{segarra2015graphfilteringTSP15}. The product and sum representations in~\eqref{eqn_diffusion} are common equivalent models for the generation of random network processes. Indeed, any process that can be understood as the linear propagation of an input signal through a static graph can be written in the form in \eqref{eqn_diffusion}, subsuming heat diffusion~\cite{vespignanibook}, consensus and the classic DeGroot model of opinion dynamics~\cite{DeGrootConsensus}, as well as symmetric structural equation models (SEMs) \cite{kolaczyk2009book} as special cases.
When $\bbx$ is white so that $\bbC_\bbx=\bbI_N$, \eqref{eqn_diffusion} is equivalent to saying that the graph process $\bby$ is \emph{stationary} in $\bbS$; see e.g., \cite[Def. 1]{marques2016stationaryTSP16},~\cite{perraudinstationary2016},~\cite{girault_stationarity} and Section \ref{ssec:stationarity} for further details. Here though, we deal with more general non-stationary signals $\bby$ that adhere to linear diffusion dynamics as in \eqref{eqn_diffusion}, but where the input covariance $\bbC_{\bbx}$ can be arbitrary. This is for instance relevant to (geographically) correlated sensor network data, urban population mobility patterns, or to models of opinion dynamics {among polarized groups.}

The justification to say that $\bbS$ represents the structure of $\bby$ is that we can think of the edges of $\bbS$ as direct (one-hop) relations between the elements of the signal. The diffusion described by \eqref{eqn_diffusion} modifies the original correlation by inducing indirect (multi-hop) relations. In this context, our goal is to recover the fundamental relations dictated by $\bbS$ from a set of independent samples of a non-stationary random signal $\bby$, as well as realizations of $\bbx$, or more pragmatically, knowledge of $\bbC_\bbx$. This additional information on the input $\bbx$ is the price paid to accommodate the more general non-stationary generative models for $\bby$, and is not needed when identifying the structure of stationary graph signals~\cite{segarra2016topoidTSP16}, since $\bbC_\bbx = \bbI_N$ in that case.

\noindent \textbf{Relation to prior work.} Under the assumption that the signals are related to the topology of the graph where they are supported,
the goal of graph signal processing (GSP) is to develop algorithms that fruitfully leverage this relational structure,
and can make inferences about these relationships when they are only partially observed~\cite{EmergingFieldGSP,SandryMouraSPG_TSP13}. Most
GSP efforts to date assume that the underlying network is known, and then analyze how the graph's algebraic
and spectral characteristics impact the properties of the graph signals of interest. However, such assumption
is often untenable in practice and arguably most graph construction schemes are largely informal, distinctly
lacking an element of validation. 

Network topology inference is a prominent task in Network Science~\cite[Ch. 7]{kolaczyk2009book},~\cite{SI_SPMAG}. 
Since networks typically encode similarities between nodes, several topology, inference approaches construct graphs whose edge weights correspond to nontrivial correlations or coherence measures between signal profiles at incident nodes~\cite{kolaczyk2009book,sporns2012book}. 
Acknowledging that the observed correlations can be due to latent network effects, alternative methods rely on inference of full partial correlations~\cite[Ch. 7.3.2]{kolaczyk2009book}. Under Gaussianity assumptions, there are well-documented connections with covariance selection~\cite{dempster_cov_selec} and sparse precision matrix estimation~\cite{GLasso2008,banerjee2008jlmr,Lake10discoveringstructure,slawski2015estimation,egilmez2017jstsp}, as well as high-dimensional sparse linear regression~\cite{meinshausen06}. Extensions to directed graphs include SEMs~\cite{BazerqueGeneNetworks,BainganaInfoNetworks,shen2017tensors}, Granger causality~\cite{Brovelli04Granger,sporns2012book}, or their nonlinear (kernelized) variants~\cite{Karanikolas_icassp16,shen2016kernelsTSP16}.  Recent GSP-based network inference frameworks postulate instead that the network exists as a latent underlying structure, and that observations are generated as a result of a network process defined in such graph~\cite{segarra2016topoidTSP16,pasdeloup2016inferenceTSIPN16,thanou17,MeiGraphStructure,DongLaplacianLearning,Kalofolias2016inference_smoothAISTATS16}. Different from~\cite{thanou17,DongLaplacianLearning,MeiGraphStructure, Kalofolias2016inference_smoothAISTATS16} that operate on the graph domain, the goal here is to identify graphs that endow the given observations with desired spectral (frequency-domain) characteristics. Two works have recently explored this approach by identifying a GSO given its eigenvectors~\cite{segarra2016topoidTSP16,pasdeloup2016inferenceTSIPN16}, but both rely on observations of stationary graph signals. Different from~\cite{DongLaplacianLearning,Kalofolias2016inference_smoothAISTATS16,sandeep_icassp17,mike_icassp17} that infer structure from signals assumed to be smooth over the sought graph, here the measurements are related to the
graph via filtering (e.g., modeling the diffusion of an idea or the spread of a disease).  Smoothness models
are subsumed as special cases found with diffusion filters having a low-pass frequency response.

\noindent \textbf{Paper outline.} In Section \ref{S:prelim_problem} we formulate the problem of identifying a GSO that explains the fundamental structure of a random signal diffused on a graph. While for stationary $\bby$ the sought GSO shares its eigenvectors with the signal's covariance matrix~\cite{marques2016stationaryTSP16, perraudinstationary2016,segarra2016topoidTSP16}, in the general (non-stationary) setting dealt with here this no longer holds and we elaborate on the ensuing challenges (Section \ref{ssec:stationarity}). Still, the graph's eigenvectors are preserved by the polynomial graph filter that governs the underlying diffusion dynamics~\eqref{eqn_diffusion}. This motivates a novel two-step network topology inference approach whereby we: i) identify the GSO's eigenbasis from a judicious \textit{graph filter estimate}; and ii) rely on these \emph{spectral templates} to estimate the GSO's eigenvalues such that the inferred graph exhibits desirable structural characteristics (e.g.,  sparsity or minimum-energy edge weights; see also Section \ref{Ss:Finding_the_eigenvalues}). The estimation of the diffusion filter in step i), which is not required when the signals are stationary~\cite{segarra2016topoidTSP16}, has merit on its own and is of interest beyond topology inference. Feasibility of this additional system identification task requires extra information on the excitation signal $\bbx$. Section \ref{S:NonStationary_InputSignals} addresses the (simpler) setup where direct observations of the inputs are available so that the unknown filter matrix and the input-output signal pairs are linearly related. The focus in Section \ref{S:NonStationary_InputStatistics} shifts to scenarios where second-order statistical information is used, and the relationship between the input-output covariances and the filter is quadratic. Identifiability conditions for the noise-free case are discussed and particular cases for which the problem can be recast as a convex quadratic optimization are described. Section \ref{S:algorithms} develops projected gradient descent and semidefinite relaxation-based algorithms with complementary strengths, to deal with the (non-convex) fourth-order polynomial minimization associated with the recovery problem in Section \ref{S:NonStationary_InputStatistics}. Numerical tests with synthetic and real-world data corroborate the effectiveness of the novel approach in recovering the topology of social, brain{, financial} and transportation networks (Section \ref{S:Simulations}).  Concluding remarks are given in Section \ref{S:Conclusions}.

\noindent \textbf{Notation.} The entries of a matrix $\mathbf{X}$ and a (column) vector $\mathbf{x}$ are denoted by $X_{ij}$ and $x_i$, respectively. Sets are represented by calligraphic capital letters. 
The notation $^T$ and $^\dag$ stands for transpose and pseudo-inverse, respectively; $\mathbf{0}$ and $\mathbf{1}$ refer to the all-zero and all-one vectors; while $\bbI_N$ denotes the $N\times N$ identity matrix. For a vector $\bbx$, $\diag(\mathbf{x})$ is a diagonal matrix whose $i$th diagonal entry is $x_i$. 
The operators $\otimes$, $\odot$, and $\text{vec}(\cdot)$ stand for Kronecker product, Khatri-Rao (columnwise Kronecker) product, and matrix vectorization, respectively. Lastly, $\| \bbX \|_p$ denotes the $\ell_p$ norm of $\text{vec}(\bbX)$ and $\mathrm{ker}(\bbX)$ refers to the null space of $\bbX$. The spectral radius of matrix $\bbX$ is denoted by $\lambda_{\max}(\bbX)$.

\section{Problem Statement}\label{S:prelim_problem}

Consider the generative model in \eqref{eqn_diffusion}, whereby the properties of the graph signal $\bby$ depend on those of the excitation input $\bbx$ and the underlying graph $\ccalG$ represented by the GSO $\bbS$.  
Given realizations of the output and prior information on the input, the goal is to infer a parsimonious graph representation that explains the structure of $\bby$. Alternatively, we can say that the goal is to recover the GSO which encodes direct relationships between the elements of $\bby$ from observable indirect relationships generated by a diffusion process. 

To formally state the problem, consider the symmetric GSO $\bbS$ associated with the undirected graph $\ccalG$. Define the eigenvector matrix $\bbV:=[\bbv_1,\ldots,\bbv_N]$ and the eigenvalue matrix $\bbLam:=\diag(\lam_1,\ldots,\lam_N)$ to write $\bbS =\bbV\bbLam\bbV^T$.
Now observe that while the diffusion expressions in \eqref{eqn_diffusion} are polynomials on the GSO of possibly infinite degree, the Cayley-Hamilton theorem asserts they are equivalent to polynomials of degree smaller than $N$. Upon defining the vector of coefficients $\bbh:=[h_0,\ldots,h_{L-1}]^T\in\reals^L$ and the symmetric graph filter $\bbH:=\sum_{l=0}^{L-1} h_l \bbS^l\in \reals^{N\times N}$~\cite{SandryMouraSPG_TSP13}, the generative model in \eqref{eqn_diffusion} can be rewritten as
\begin{equation}\label{E:Filter_input_output_time}
	\bby  = \textstyle \big(\sum_{l=0}^{L-1}h_l \bbS^l\big)\,\bbx
	= \bbH \bbx, 
\end{equation}
for some particular $\bbh$ and $L\leq N$. Given a set $\ccalY\!:=\!\{\bby^{(p)}\}_{p=1}^P$ of $P$ independent samples of a non-stationary random signal $\bby$ adhering to the network diffusion model \eqref{E:Filter_input_output_time}, the problem is to identify the GSO $\bbS$ which is optimal in some sense as described in Section \ref{Ss:Finding_the_eigenvalues}. 

Fundamental to the topology inference approach developed here is to note that because $\bbH$ is a polynomial on $\bbS$, then: i) all such graph filters (spanned by the unknown coefficients $\bbh$) have the \textit{same eigenvectors}; and ii) such eigenvectors are the same as those of the shift, namely $\bbV$. In other words, while the diffusion implicit in $\bbH$ obscures the eigenvalues of the GSO, the eigenvectors $\bbV$ are preserved as \emph{spectral templates} of the underlying network topology. Next, Section \ref{ssec:stationarity} describes how to leverage \eqref{E:Filter_input_output_time} to obtain the GSO eigenbasis from a set of nodal observations $\ccalY$, by first estimating the unknown graph filter $\bbH$. We show that the information in $\ccalY$ is in general not enough to uniquely recover $\bbH$. Hence, we will resort to additional knowledge on the input signal $\bbx$ (either realizations, sparsity properties, or, second-order statistical information) and also possibly on the structure of the graph filter $\bbH$. Section \ref{Ss:Finding_the_eigenvalues} outlines how to use the spectral templates $\bbV$ to recover the desired GSO by estimating its eigenvalues $\bbLambda$ and, as byproduct, the graph shift $\bbS=\bbV\bbLambda\bbV^T$ itself. 

\subsection{Stationary versus non-stationary observations}\label{ssec:stationarity}
Consider that the $P$ observations in $\ccalY$ correspond to independent realizations of a process $\bby$ adhering to the generative model in \eqref{E:Filter_input_output_time}. The goal is to use $\ccalY$ to estimate the spectral templates $\bbV$ of the filter $\bbH$ that governs the diffusion in \eqref{E:Filter_input_output_time}.

To gain insights, suppose first that $\bbx$ is white so that $\bbC_\bbx = \bbI_N$~\cite{segarra2016topoidTSP16}. Then  
the covariance matrix of $\bby=\bbH\bbx$ is
\begin{equation}\label{E:covariance_y}
\bbC_\bby:=\EE[\bby\bby^T]=\EE[\bbH\bbx(\bbH\bbx)^T]= \bbH \EE[\bbx\bbx^T]\bbH	= \bbH^2.
\end{equation}
In obtaining the third equality {we} used that $\bbH$ is symmetric, because it is a polynomial in the symmetric GSO $\bbS$. Using the spectral decomposition of $\bbS =\bbV\bbLam\bbV^T$ to express the filter as $\bbH = \sum_{l=0}^{L-1} h_l (\bbV\bbLam\bbV^T)^l =\bbV(\sum_{l=0}^{L-1} h_l \bbLam^l)\bbV^T$, we can diagonalize the covariance matrix as 
\begin{equation}\label{eqn_diagonalize_covariance}
\bbC_\bby=\textstyle  \bbV\left(\sum_{l=0}^{L-1}h_l\bbLam^l\right)^2\bbV^T.
\end{equation}
Such a covariance expression is precisely the requirement for a graph signal to be stationary in $\bbS$~\cite[Def. 2.b]{marques2016stationaryTSP16}. Remarkably, if $\bby$ is graph stationary, or equivalently if $\bbx$ is white, \eqref{eqn_diagonalize_covariance} shows that the \textit{eigenvectors} of the shift $\bbS$, the  filter $\bbH$, and the covariance $\bbC_\bby$ are \textit{all the same}. 
As a result, to estimate $\bbV$ from the observations $\{\bby^{(p)}\}_{p=1}^P$ it suffices to form the \textit{sample covariance} $\hbC_{\bby}=\frac{1}{P}\sum_{p=1}^{P}\bby^{(p)}(\bby^{(p)})^T$ 
and use its eigenvectors as spectral templates to recover $\bbS$~\cite{segarra2016topoidTSP16,pasdeloup2016inferenceTSIPN16}; see also Section \ref{Ss:Finding_the_eigenvalues}.  {Note that in estimating $\bbC_{\bby}$, we assume that the observed signals are zero-mean without loss of generality, otherwise we can subtract the mean from the signals.}

\begin{figure}[t]
	\begin{minipage}[b]{.5\linewidth}
		\centering
		\includegraphics[width=\linewidth, trim={0cm, 0cm, 0cm, 1cm}]{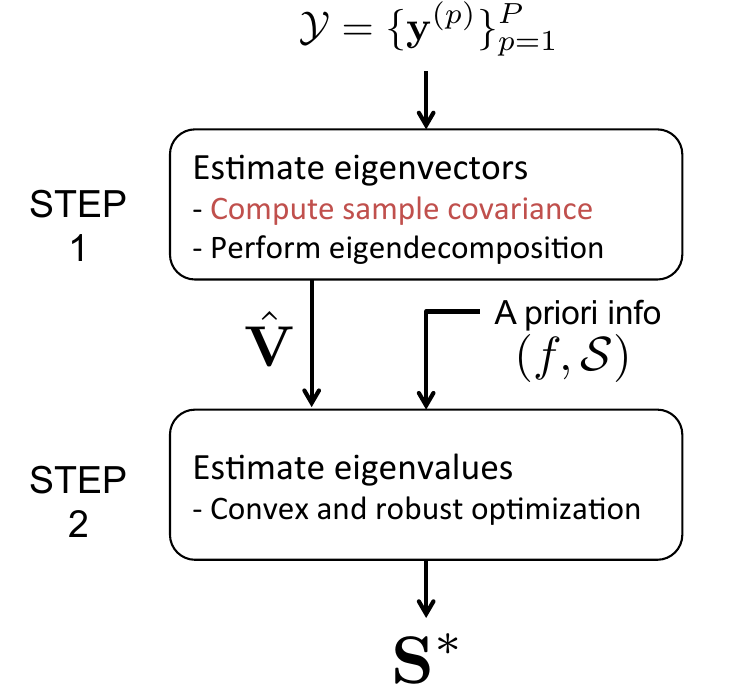}
	\end{minipage}
	%
	%
	\begin{minipage}[b]{.48\linewidth}
		\centering
		\includegraphics[width=\linewidth, trim={0cm, 0cm, 0cm, 1cm}]{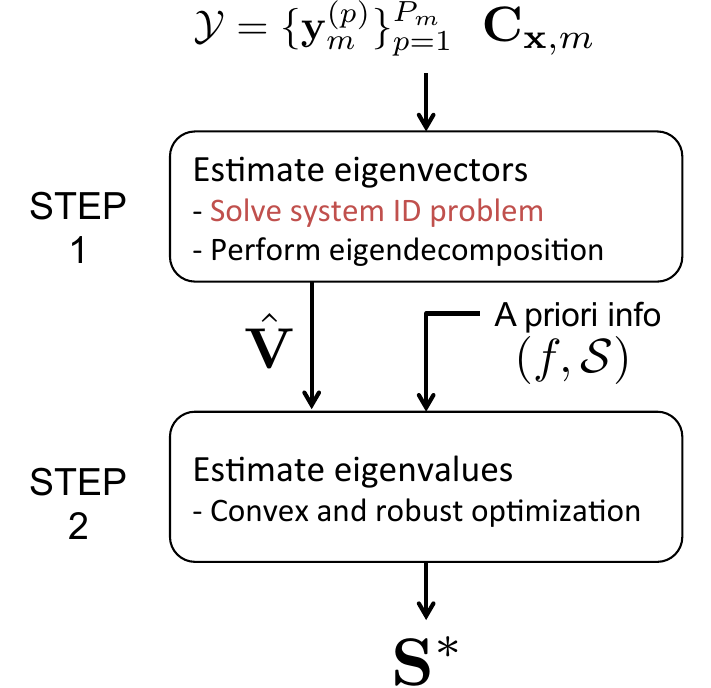}
	\end{minipage}
	%
	\vspace*{-0.2cm}
	\caption{{Schematic view of the two-step network inference method for (left) stationary and (right) non-stationary observations from diffusion processes. The main differences between both approaches lie in Step 1. 
		For non-stationary processes, covariance matrices are no longer simultaneously diagonalizable with $\bbS$, thus requiring a more challenging system identification step in order to estimate $\bbH$ (see Sections \ref{S:NonStationary_InputSignals}-\ref{S:algorithms}).
		In both cases the output of Step 1 is an estimate $\hbV$ of the eigenvectors of the sought shift. During Step 2, this estimate is combined with a priori information about the shift in an optimization problem to obtain the estimate $\bbS^*$, as described in Section \ref{Ss:Finding_the_eigenvalues}.}}
	\label{F:scheme_topo_id}
	
\end{figure}
%

In this context, the broader focus of the present paper is on identifying the GSO $\bbS$ that is considered to be the best possible description of the structure of a \emph{non-stationary} signal $\bby=\bbH\bbx$ [cf. \eqref{E:Filter_input_output_time}, where $\bbx$ is not white]. For generic (non-identity) input covariance matrix $\bbC_\bbx$, we face the challenge that the signal covariance [cf. \eqref{E:covariance_y}]
\begin{equation}\label{E:covariance_y_ns}
\bbC_\bby=\bbH\bbC_\bbx\bbH
\end{equation}
\emph{is no longer simultaneously diagonalizable with $\bbS$}. This rules out
using the eigenvectors of the sample covariance $\hbC_\bby$ as spectral templates of $\bbS$. Still, as argued following \eqref{E:Filter_input_output_time} the eigenvectors of the GSO coincide with those of the graph filter $\bbH$ that governs the underlying diffusion dynamics. This motivates using realizations of observed signals together with additional information on the excitation inputs $\bbx$ (either realizations of the graph signals, sparsity assumptions, or the covariance matrix $\bbC_{\bbx}$ \cite{shen2017tensors}) to \textit{identify the filter} $\bbH$, with the ultimate goal of estimating its eigenvectors $\bbV$. This system identification task in the graph setting is the subject dealt with in Sections \ref{S:NonStationary_InputSignals} and \ref{S:NonStationary_InputStatistics},
{but before moving on, we close} the loop {by} showing how to recover $\bbS$ given its estimated eigenbasis $\hbV${; see also the comparative schematic in Fig. \ref{F:scheme_topo_id}.} 

\subsection{Using the spectral templates to recover the shift}\label{Ss:Finding_the_eigenvalues}
Given estimates $\hbV$ of the filter eigenvectors, recovery of $\bbS$ amounts to selecting its eigenvalues $\bbLam$ and to that end we assume that the shift of interest is optimal in some sense. At the same time, we should account for the discrepancies between $\hbV$ and the actual eigenvectors of $\bbS$, due to finite sample size constraints and unavoidable errors in estimating the filter $\bbH$. Accordingly, we build on~\cite{segarra2016topoidTSP16} and seek for the shift operator $\bbS$ that: (a)  is optimal with respect to (often convex) criteria $f ( \bbS)$; (b) belongs to a convex set $\ccalS$ that specifies the desired type of shift operator (e.g., the adjacency $\bbA$ or Laplacian $\bbL$); and (c) is close to $\hbV\bbLambda\hbV^T$ as measured by a convex matrix distance $d(\cdot,\cdot)$. Formally, one can solve
\begin{equation}\label{E:general_problem}
\bbS^* := \argmin_{\bbLambda, \bbS \in \ccalS} \
f ( \bbS) ,   \quad
\text{s. to }\:d(\bbS,\hbV\bbLambda\hbV^T)\leq \epsilon
\end{equation}
which is a convex optimization problem provided $f ( \bbS)$ is convex, and $\epsilon$ is a tuning parameter chosen based on a priori information on the imperfections. 
Within the scope of the signal model \eqref{eqn_diffusion}, the formulation \eqref{E:general_problem} entails a general class of network topology inference problems parametrized by the choices in (a)-(c) above. The spectrum of alternatives is briefly outlined next, while concrete choices are made for the numerical tests in Section \ref{S:Simulations}.

\noindent\textbf{Criteria.} The selection of  $f(\bbS)$ allows to incorporate physical characteristics of the desired graph into the formulation, while being consistent with the spectral templates $\hbV$. For instance, the matrix (pseudo-)norm $f(\bbS)=\|\bbS\|_0$ which counts the number of nonzero entries in $\bbS$ can be used to minimize the number of edges towards identifying sparse graphs (e.g., of direct relations among signal elements); $f ( \bbS)=\|\bbS\|_1$ is a convex proxy for the aforementioned edge cardinality function.  Alternatively, the Frobenius norm $ f ( \bbS)=\|\bbS\|_F$ can be adopted to minimize the energy of the edges in the graph, or $f ( \bbS)=\|\bbS\|_{\infty}$ can be chosen to obtain shifts $\bbS$ associated with graphs of uniformly low edge weights. This can be meaningful when identifying graphs subject to capacity constraints.

\noindent\textbf{Constraints.} The constraint $\bbS \in \ccalS$ in \eqref{E:general_problem} incorporates a priori knowledge about $\bbS$. If we let $\bbS = \bbA$ represent the adjacency matrix of an undirected graph with non-negative weights and no self-loops, we can explicitly write $\ccalS$ as follows
\begin{align} \label{E:SparseAdj_def_S}
\ccalS_{\mathrm{A}} \!:= \! \{ \bbS \, | \, S_{ij} \geq 0, \;\,  \bbS\!\in\!\ccalH_N\!,\;\,  S_{ii} = 0, \;\, \textstyle\sum_j S_{j1} \! = \! 1 \}.
\end{align}
The first condition in $\ccalS_{\mathrm{A}}$ encodes the non-negativity of the weights whereas the second condition incorporates that $\ccalG$ is undirected, hence, $\bbS$ must belong to the set $\ccalH_N$ of real and symmetric $N \! \times \! N$ matrices. The third condition encodes the absence of self-loops, thus, each diagonal entry of $\bbS$ must be null. Finally, the last condition fixes the scale of the admissible graphs by setting the weighted degree of the first node to $1$, and rules out the solution $\bbS\!=\!\bbzero$. Other GSOs (e.g., the Laplacian $\bbL$ and its normalized variants) can be accommodated in our framework via minor modifications to $\ccalS$; see~\cite{segarra2016topoidTSP16}. 

The form of the convex matrix distance $d(\cdot, \cdot)$ depends on the particular application. For instance, if $\|\bbS-\hbV\bbLambda\hbV^T\|_F$ is chosen the focus is more on the similarities across the entries of the shifts, while $\|\bbS-\hbV\bbLambda\hbV^T\|_2$ focuses on their spectrum.

\section{(Bi)linear graph filter identification}\label{S:NonStationary_InputSignals}

Consider $m=1,\ldots,M$ diffusion processes on $\ccalG$, and assume that the observed non-stationary signal $\bby_m$ corresponds to an input $\bbx_m$ diffused by an unknown graph filter $\bbH=\sum_{l=0}^{L-1} h_l\bbS^l$, which encodes the structure of the network via $\bbS$. In this section we show how additional knowledge about \emph{realizations} of the input signals $\bbx_m$ can be used to identify $\bbH$ and, as byproduct, its eigenvectors $\bbV$. We consider settings in which this extra information comes either from direct observation of $\{\bbx_m\}_{m=1}^M$, or through an assumption on input signal sparsity. {In the context of online media-based marketing campaigns or rumor (fake news) diffusion over social networks, the observable input graph signal could correspond to the initial excitation instilled by those known (often paid) influencers. In neuroscience, the observed inputs  may represent controlled external stimuli aimed at exciting a few neural regions via e.g., transcranial magnetic stimulation.}

\subsection{Input-output signal realization pairs}\label{Ss:linear_equations}

Suppose first that realizations of $M$ output-input pairs $\{\bby_m,\bbx_m\}_{m=1}^M$ are available, which can be arranged in the data matrices $\bbY=[\bby_1,...,\bby_M]$ and $\bbX=[\bbx_1,...,\bbx_M]$. The goal is to identify a symmetric filter $\bbH\in\ccalH_N$ such that the observed signal $\bby_m$ and the predicted one $\bbH \bbx_m$ are close in some sense. In the absence of measurement noise this simply amounts to solving a system of $M$ linear matrix equations 
\begin{equation}\label{E:linear_system}
	\bby_m=\bbH\bbx_m, \quad m=1,\ldots,M.
\end{equation}
When noise is present, using the workhorse least-squares (LS) criterion the filter can be estimated as 
\begin{equation}\label{E:opt_filter_inputoutput_opt}
	\bbH^*\!\!=\!\mathop{\mathrm{argmin}}_{\bbH\in\ccalH_N }\sum_{m=1}^M\!\!\|\bby_m\!-\!\bbH\bbx_m\|_2^2. 
\end{equation}
Because $\bbH$ is symmetric, the free optimization variables in \eqref{E:opt_filter_inputoutput_opt} correspond to, say, the lower triangular part of $\bbH$, meaning the entries on and below the main diagonal. These $N_{\bbH}:=N(N\!+\!1)/2$ non-redundant entries can be conveniently arranged in the so-termed half-vectorization of $\bbH$, i.e., a vector $\textrm{vech}(\bbH) \in \reals^{N_{\bbH}}$ from which one can recover $\textrm{vec}(\bbH)\in\reals^{N^2}$ via duplication. Indeed, there exists a unique duplication matrix $\bbD_N \in \{0,1\}^{N^2 \times {N_{\bbH}}}$ such that one can write $\bbD_N \text{vech}(\bbH) = \text{vec}({\bbH})$~\cite{duplication_matrices}. The Moore-–Penrose pseudoinverse of $\bbD_N$, denoted as $\bbD_N^\dagger$, possesses the property $\text{vech}(\bbH) =\bbD_N ^\dagger \text{vec}({\bbH})$.  With this notation in place, several properties of the solution $\bbH^*$ of \eqref{E:opt_filter_inputoutput_opt} are stated next.

\begin{myproposition}\label{P:Closedform_inputoutputpairs}
	Let $M_r$ denote the rank of matrix $\bbX\in\reals^{N\times M}$. Then, it holds that:\\
	\noindent a) The entries of the symmetric filter $\bbH^*$ that solves \eqref{E:opt_filter_inputoutput_opt} are  
	\begin{equation}
		\normalfont\text{vech}(\bbH^*) = \left[\big(\bbX^T\otimes\bbI_{N}\big)\bbD_N\right]^\dagger\text{vec}(\bbY).\label{E:opt_filter_inputoutput_closedform}
	\end{equation}
	\noindent b) ${\normalfont\text{rank}}\left(\big(\bbX^T\otimes\bbI_{N}\big)\bbD_N\right)\leq N_{\bbH}-(N-M_r+1)(N-M_r)/2$.\\
	\noindent c) The minimizer of \eqref{E:opt_filter_inputoutput_opt} is unique if and only if $M_r\! =\! N$.
\end{myproposition}

\noindent \textbf{Proof:}
See Appendix~\ref{ProofAppMinimizerInputOutput}. \hfill $\blacksquare$

\noindent Proposition \ref{P:Closedform_inputoutputpairs} asserts that if the excitation input set $\{\bbx_m\}_{m=1}^M$ is sufficiently rich -- i.e., if $M\geq N$ and the excitation signals are linearly independent --, the entries of the diffusion filter $\bbH$ can be found as the solution of an LS problem. Interestingly, {the fact} that $\bbH$ has only $N(N+1)/2$ different entries cannot be exploited to reduce the number $M$ of input signals required to identify the filter. The reason is that the matrix $(\bbX^T \otimes \bbI_N)\bbD_N$ is rank deficient if $\bbX^T$ has a non-trivial  null space. In other words, when using input-output pairs to estimate the filter $\bbH$ one needs the same number of pairs, regardless of whether the graph is symmetric or not. Symmetry, however, can be exploited to enhance performance in ovetermined scenarios with noisy observations; see also the tests in Section \ref{S:Simulations}. 

As explained in Section \ref{Ss:Finding_the_eigenvalues}, once $\bbH^*$ is estimated using \eqref{E:opt_filter_inputoutput_closedform}, the next step is to decompose the filter as $\bbH^*=\hbV\hbLambda\hbV^T$ and use $\hbV$ as input for the GSO identification problem \eqref{E:general_problem}. Note that obtaining such an eigendecomposition is always possible since filter estimates $\bbH^*\in\ccalH_N$ are constrained to be symmetric.

\subsection{Sparse input signals}\label{Ss:input_sparsity}

It is not uncommon to encounter application domains in which the diffused graph signals adhere to linear network dynamics $\bby=\bbH\bbx$ and the input $\bbx$ is sparse, having only a few nonzero entries. Sparsity in $\bbx$ is well-motivated due to its practical relevance and modeling value -- network processes such as $\bby$ are oftentimes the diffused version of few
localized sources, hereby indexed by $\text{supp}(\bbx):=\{i:x_{i}\neq 0\}$~\cite{segarra2016blind,thanou17}. For instance, opinion formation processes in social networks have been modeled using graph filters (see e.g.,~\cite{segarra2015graphfilteringTSP15, DeGrootConsensus}), and sparse $\bbx$ could represent the initial opinion of those few influential actors that instilled the observed status-quo. Similar ideas are naturally relevant to linear network dynamics encountered with rumor spreading, adoption of new technologies, epidemic outbreaks~\cite{segarra2016blind}, as well as with identification of heat, pollutant, or seismic localized sources~\cite{thanou17}. 

Given realizations of $M$ diffusion processes $\{\bby_m\}_{m=1}^M$ arranged as columns of matrix $\bbY\in\reals^{N\times M}$, a possible formulation of the graph filter identification problem amounts to finding $\bbH\in\ccalH_N$ such that $\bbY$ is close to $\bbH\bbX$, where the unobserved matrix $\bbX=[\bbx_1,...,\bbx_M]$ is assumed to be sparse. Different from Section \ref{Ss:linear_equations}, input realizations are now unavailable and the resulting bilinear problem entails finding the decomposition factors up to unavoidable scaling and permutation ambiguities. In the absence of noise, recent fundamental results and accompanying algorithms developed in~\cite{pmlr-v23-spielman12} can be brought to bear here. Regarding identifiability, it is established therein that $M=\ccalO(N \log N)$ samples are sufficient to uniquely determine the
decomposition with high probability, under the assumption that $\bbX$ is generated by a sparsity-inducing Bernoulli-Gaussian
or Bernoulli-Rademacher process~\cite[Theorem 1]{pmlr-v23-spielman12}. From an algorithmic standpoint, an efficient dictionary learning procedure called Exact Recovery of Sparsely-Used Dictionaries (ER-SpUD) is proposed that solves a sequence
of linear programs with varying constraints. Under the aforementioned assumptions,
the algorithm exactly recovers $\bbH$ and $\bbX$ with high probability. This result holds when: (i) the sparsity level measured by the expected
number of nonzero elements in each column of $\bbX$ is at most of order $\sqrt{N}$; and (ii) the number of samples $M$ is
at least of order $N^2\log^2N$.



\section{Quadratic graph filter identification}\label{S:NonStationary_InputStatistics}

In a number of applications, realizations of the excitation input process $\bbx_m$ may be challenging to acquire, but information about the \textit{statistical} description of $\bbx_m$ could still be available. To be specific, assume that the excitation input processes are zero mean and their covariance $\bbC_{\bbx,m}=\EE[\bbx_m\bbx_m^T]$ is known. Further suppose that for each input $\bbx_m$, we have access to a set of output observations $\{\bby_m^{(p)}\}_{p=1}^{P_m}$, which are then used to estimate the output covariance as $\hbC_{\bby,m}=\frac{1}{P_m}\sum_{p=1}^{P_m}\bby_m^{(p)}(\bby_m^{(p)})^T$. Since under \eqref{E:Filter_input_output_time} the \textit{ensemble covariance} is $\bbC_{\bby,m} = \EE [\bby_m\bby_m^T] = \bbH \bbC_{\bbx,m}\bbH$ [cf. \eqref{E:covariance_y_ns}], the aim is to identify a filter $\bbH$ such that matrices $\hbC_{\bby,m}$ and $\bbH \bbC_{\bbx,m} \bbH$ are close.


Assuming for now perfect knowledge of the signal covariances, the above rationale suggests studying the solutions of the system of matrix \emph{quadratic} equations
\begin{equation}\label{E:quadratic_system}
\bbC_{\bby,m}=\bbH \bbC_{\bbx,m} \bbH, \quad m=1,\ldots,M.
\end{equation}
To gain some initial insights, consider first the case where $M=1$ and henceforth drop the subindex $m$ so that we can write \eqref{E:quadratic_system} as \eqref{E:covariance_y_ns}. Given the eigendecomposition of the symmetric and positive semidefinite (PSD) covariance matrix $\bbC_{\bby}=\bbV_{\bby}\bbLambda_{\bby}\bbV_{\bby}^T$, the \emph{principal square root} of $\bbC_{\bby}$ is the unique symmetric and PSD matrix $\bbC_{\bby}^{1/2}$ which satisfies $\bbC_{\bby}=\bbC_{\bby}^{1/2}\bbC_{\bby}^{1/2}$. It is given by
$\bbC_{\bby}^{1/2}=\bbV_{\bby}\bbLambda_{\bby}^{1/2}\bbV_{\bby}^T$, where $\bbLambda_{\bby}^{1/2}$ stands for a diagonal matrix with the nonnegative square roots of the eigenvalues of $\bbC_{\bby}$. 

With this notation in place,  introduce the matrices $\bbC_{\bbx\bby\bbx}:=\bbC_{\bbx}^{1/2} \bbC_{\bby} \bbC_{\bbx}^{1/2}$ and $\bbH_{\bbx\bbx}:=\bbC_{\bbx}^{1/2}\bbH \bbC_{\bbx}^{1/2}$. Clearly, $\bbC_{\bbx\bby\bbx}$ is both symmetric and PSD. Regarding the transformed filter $\bbH_{\bbx\bbx}$, note that by construction we have that $\bbH_{\bbx\bbx}$ is symmetric. Moreover, if $\bbH$ is assumed to be PSD, then so will be $\bbH_{\bbx\bbx}$. These properties will be instrumental towards characterizing the solutions of the matrix quadratic equation $\bbC_\bby=\bbH\bbC_\bbx\bbH$ in \eqref{E:covariance_y_ns}, which can be equivalently recovered from the solutions $\bbH_{\bbx\bbx}$ of 
\begin{equation}\label{E:square_root_NS_one_sym}
\bbC_{\bbx\bby\bbx}\!=\!\bbC_{\bbx}^{1/2} \bbC_{\bby} \bbC_{\bbx}^{1/2}\!=\!\bbC_{\bbx}^{1/2}\bbH \bbC_{\bbx} \bbH \bbC_{\bbx}^{1/2}\!=\!\bbH_{\bbx\bbx}^2. 
\end{equation}
{This relationship has the same quadratic form as its counterpart for the stationary case [cf. \eqref{E:covariance_y}], with the identifications $\bbC_{\bbx\bby\bbx} \leftrightarrow \bbC_{\bby}$ and $\bbH_{\bbx\bbx} \leftrightarrow \bbH$. However, there is no apparent way to relate the eigenvectors of $\bbH_{\bbx\bbx}$ with those of $\bbH$, namely the desired spectral templates $\bbV$. Consequently, the problem does not boil down to the white case in Section \ref{ssec:stationarity}.}

{
\begin{remark}[Input realizations versus covariances] \label{R:Linear_vs_Quadract}\normalfont
If input signals are available as in Section \ref{Ss:linear_equations}, these  could be used to estimate input covariances. However, tackling the problem when input-output realization pairs are given entails solving a linear matrix equation \eqref{E:linear_system}. On the other hand, filter identification based on covariance information requires solving a system of quadratic matrix equations \eqref{E:quadratic_system}, which is more challenging. Moreover, estimating the input and output covariances a priori typically requires a larger sample relative to the one needed to solve the linear system identification task directly.
\end{remark}
}

\subsection{Positive semidefinite graph filters}\label{Ss:PSD_filter}

Let us suppose first that $\bbH$ is PSD (henceforth denoted $\bbH\in\ccalH_N^{++}$), so that $\bbH_{\bbx\bbx}$ in \eqref{E:square_root_NS_one_sym} is PSD as well. Such filters arise, for example, with heat diffusion processes of the form $\bby=(\sum_{l=0}^\infty \beta^l \bbL^l)\bbx$ with $\beta>0$, where the Laplacian shift $\bbL$ is PSD and the filter coefficients $h_l=\beta^l$ are all positive. In this setting, the solution of  \eqref{E:square_root_NS_one_sym} is unique and given by the principal square root
\begin{equation}
\bbH_{\bbx\bbx}=\bbC_{\bbx\bby\bbx}^{1/2}.
\end{equation}
Consequently, if $\bbC_{\bbx}$ is nonsingular (so that the excitation inputs are not degenerate), the definition of $\bbH_{\bbx\bbx}$ can be used to recover $\bbH$ via 
\begin{equation}\label{E:filt_estimate_psd}
\bbH = \bbC_{\bbx}^{-1/2} \bbC_{\bbx\bby\bbx}^{1/2} \bbC_{\bbx}^{-1/2}.
\end{equation} 
The previous arguments demonstrate that the assumption $\bbH\in\ccalH_N^{++}$ gives rise to a strong identifiability result. Indeed, if $\{\bbC_{\bby,m}\}_{m=1}^M$ are known perfectly, a PSD graph filter is identifiable even for $M=1$. 

However, in pragmatic settings where only empirical covariances are available, then observation of multiple ($M>1$) diffusion processes can improve the performance of the system identification task. Given empirical covariances $\{\hbC_{\bby,m}\}_{m=1}^M$  respectively estimated with enough samples $P_m$ to ensure they are of full rank, for each $m$ define $\hbC_{\bbx\bby\bbx,m}:=\bbC_{\bbx,m}^{1/2} \hbC_{\bby,m} \bbC_{\bbx,m}^{1/2}$. The quadratic equation \eqref{E:square_root_NS_one_sym} motivates solving the LS problem
\begin{equation}\label{E:opt_filter_PSD_opt}
\bbH^*=\mathop{\mathrm{argmin}}_{\bbH\in\ccalH_N^{++}}\sum_{m=1}^M\|\hbC_{\bbx\bby\bbx,m}^{1/2}-\bbC_{\bbx,m}^{1/2}\bbH\bbC_{\bbx,m}^{1/2}\|_F^2.
\end{equation}
Whenever the number of samples $P_m$ -- and accordingly the accuracy of the empirical covariances $\hbC_{\bby,m}$ -- differs significantly across diffusion processes $m=1,\ldots, M$, it may be prudent to introduce non-uniform coefficients to downweigh those residuals in \eqref{E:opt_filter_PSD_opt} with inaccurate covariance estimates.  The following proposition offers insights on the solution to \eqref{E:opt_filter_PSD_opt}, and extensions to weighted LS criteria are straightforward.
\begin{myproposition}\label{P:PSD_filter_opt}
Define the matrices $\barbX=[\bbC_{\bbx,1}^{1/2}\otimes\bbC_{\bbx,1}^{1/2},...,\bbC_{\bbx,M}^{1/2}\otimes\bbC_{\bbx,M}^{1/2}]^T$ and $\barbY=[\hbC_{\bbx\bby\bbx,1}^{1/2},...,\hbC_{\bbx\bby\bbx,M}^{1/2}]^T$. 
Then, the filter $\bbH^*$ that solves \eqref{E:opt_filter_PSD_opt} can be found as
\begin{equation}
\normalfont\text{vec}(\bbH^*)=\barbX^{\dagger}\text{vec}(\barbY^T).\label{E:minimizer_PSD_closedform}
\end{equation}
Moreover, if $M=1$ and matrix $\bbC_{\bbx,1}$ is nonsingular, the minimizer $\bbH^*$ is unique and given by
\begin{equation}
\bbH^*=\bbC_{\bbx,1}^{-1/2}\hbC_{\bbx\bby\bbx,1}^{1/2}\bbC_{\bbx,1}^{-1/2}.\label{E:minimizer_PSD_filter_2}
\end{equation}	
\end{myproposition}
\noindent \textbf{Proof:}
To show \eqref{E:minimizer_PSD_closedform} one can follow steps similar to those for \eqref{E:opt_filter_inputoutput_closedform} in Proposition \ref{P:Closedform_inputoutputpairs}. The identifiability result for $M=1$ follows from the arguments leading to \eqref{E:filt_estimate_psd}, noting that the cost in \eqref{P:PSD_filter_opt} vanishes if $\bbH^*$ is selected to satisfy $\hbC_{\bbx\bby\bbx,1}^{1/2}=\bbC_{\bbx,1}^{1/2}\bbH^* \bbC_{\bbx,1}^{1/2}$. Left and right multiplying both sides of the equality by $\bbC_{\bbx,1}^{-1/2}$, \eqref{E:minimizer_PSD_filter_2} follows.
 \hfill $\blacksquare$
 
\subsection{Symmetric graph filters}\label{Ss:symmetric_filter}

Consider now a more general setting whereby $\bbH$ is only assumed to be symmetric, and once more let $M=1$ first to simplify notation. With the unitary matrix $\bbV_{\mathbf{xyx}}$ denoting the eigenvectors of $\bbC_{\mathbf{xyx}}$ and with $\bbb\in\{-1,1\}^N$ being a binary (signed) vector, one can conclude that solutions to \eqref{E:square_root_NS_one_sym} have the general form 
\begin{equation}
\bbH_{\mathbf{xx}}= \bbC_{\mathbf{xyx}}^{1/2}\bbV_{\mathbf{xyx}}\diag(\bbb)\bbV_{\mathbf{xyx}}^T.
\end{equation}
If the input covariance matrix $\bbC_{\bbx}$ is nonsingular, all symmetric solutions $\bbH\in\ccalH_{N}$ of \eqref{E:square_root_NS_one_sym} [and hence of \eqref{E:covariance_y_ns}] are given by
\begin{equation}\label{E:filter_solutions_symmetric}
\bbH= \bbC_{\bbx}^{-1/2}\bbC_{\mathbf{xyx}}^{1/2}\bbV_{\mathbf{xyx}}\diag(\bbb)\bbV_{\mathbf{xyx}}^T\bbC_{\bbx}^{-1/2}.
\end{equation}
In the absence of the PSD assumption, the problem for $M=1$ is non-identifiable. Inspection of \eqref{E:filter_solutions_symmetric} shows there are $2^N$ possible solutions to the quadratic equation \eqref{E:covariance_y_ns}, which are parametrized by the binary vector $\bbb$. For the PSD setting in Section \ref{Ss:PSD_filter} the solution is unique and corresponds to $\bbb=\mathbf{1}$.

For $M>1$, the set of feasible solutions to the system of equations \eqref{E:quadratic_system} is naturally given by
\begin{align}\label{E:solution_set}
\ccalH_{M}^{\text{sym}}\!\!=&\bigcap_{m=1}^M\left\{\bbH\in\ccalH_{N} \:| \:\bbb_m\!\in\!\{-1,1\}^N\;\;\text{and}\;\;\right.\\
&\left.\bbH\!=\!\bbC_{\bbx,m}^{-1/2}\bbC_{\mathbf{xyx},m}^{1/2}\bbV_{\mathbf{xyx},m}\diag(\bbb_m)\bbV_{\mathbf{xyx},m}^T\bbC_{\bbx,m}^{-1/2}\right\}.\nonumber
\end{align}
%
It is thus conceivable that as $M$ grows and, therefore, the number of equations increases, the cardinality of $\ccalH_{M}^{\text{sym}}$ shrinks and the problem is rendered identifiable (up to an unavoidable sign ambiguity because if $\bbH\in\ccalH_{N}$ is a solution of \eqref{E:quadratic_system}, so is $-\bbH$).
%
%
Next, we show that even with two excitation inputs having covariances $\bbC_{\bbx,1}$ and $\bbC_{\bbx,2}$ with \textit{identical} eigenvectors, uniqueness can be established as long as their eigenvalues are sufficiently different.

\begin{myproposition}\label{P:identifiability}
Consider the system of quadratic equations \eqref{E:quadratic_system} for $M=2$ and suppose $\bbC_{\bbx,1}= \bbU{\normalfont\text{diag}}(\bblambda_1)\bbU^T$ and $\bbC_{\bbx,2}=\bbU{\normalfont\text{diag}}(\bblambda_2)\bbU^T$. Then \eqref{E:quadratic_system} has a unique symmetric solution, i.e., $\bbH=\bbV\bbLambda\bbV^T$ is identifiable up to a sign ambiguity if the following conditions hold:
\\
A-1) All eigenvalues in $\bblambda_1$ are distinct;  \\  
A-2) $\lambda_{1,i} \, \lambda_{2,j} \neq \lambda_{1,j} \, \lambda_{2,i}$ for all $i,j$;\\
A-3) $\bbV$ and $\bbU$ do not share any eigenvector; and\\
A-4) ${\normalfont\text{rank}}(\bbH)=N$.
\end{myproposition}
\noindent \textbf{Proof:}
See Appendix~\ref{ProofAppIdentifiabilityConvariances}.	
 \hfill $\blacksquare$
 
\noindent Conditions A-1) and A-2) encode a notion of richness on the excitation signals. In fact, condition A-2) is the specification for $M=2$ of a generalizable requirement based on the Kruskal rank of a matrix related to the eigenvalues of the excitation processes (see Appendix~\ref{ProofAppIdentifiabilityConvariances}). Under this perspective, it becomes immediate that larger $M$ facilitate the fulfillment of this more general requirement, leading to the expected conclusion that the more input processes we consider, the easier it becomes to identify $\bbH$.
Moreover, from the proof arguments it follows that symmetry of $\bbH$ is essential (see Lemma \ref{L:appendix} in Appendix~\ref{ProofAppIdentifiabilityConvariances}). Actually, if one lifts the symmetry assumption and all input covariances have the same eigenvectors, the problem remains non-identifiable even for high values of $M$ (regardless of the input covariance eigenvalues). 

\section{Algorithms}\label{S:algorithms}


Building on the findings in Section \ref{Ss:symmetric_filter}, here we propose two algorithms with complementary strengths to tackle the quadratic filter identification problem when the only assumption is for $\bbH$ to be symmetric (undirected graph), but not (necessarily) PSD.

\subsection{Projected gradient descent}\label{Ss:PGD}

Going back to the beginning of Section \ref{S:NonStationary_InputStatistics}, given realizations $\{\bby_m^{(p)}\}_{p=1}^{P_m}$ of the diffusion processes the goal is to identify a symmetric filter $\bbH\in\ccalH_N$ that drives $\{\bbH\bbC_{\bbx,m}\bbH\}_{m=1}^M$ close to the empirical covariances $\{\hbC_{\bby,m}\}_{m=1}^M$. Such quadratic functions of $\bbH$ can be formed under perfect knowledge on the input covariances $\{\bbC_{\bbx,m}\}_{m=1}^M$. 

Accordingly, adopting a constrained LS criterion yields a graph filter estimate 
\begin{equation}\label{E:opt_filter_genericfilter_opt}
\bbH^*=\mathop{\mathrm{argmin}}_{\bbH\in\ccalH_N }\! \sum_{m=1}^{M} \|\hbC_{\bby,m}-\bbH\bbC_{\bbx,m}\bbH^T\|_F^2.
\end{equation}
%
Weighted variants of the criterion could also be pertinent here, as discussed following \eqref{E:opt_filter_PSD_opt}.
Problem \eqref{E:opt_filter_genericfilter_opt} is a non-convex fourth-order polynomial optimization, which can potentially have multiple solutions. Since finding $\bbH^*$ is challenging, we seek algorithms capable of finding stationary solutions. 

{A viable approach is to rely on projected gradient descent (PGD)~\cite{bertsekas_nonlinear}, which 
boils down to the 
provably convergent iterations tabulated under Algorithm \ref{A:PGD}~\cite[Prop. 2.3.1]{bertsekas_nonlinear}. 
The updates entail multiplications and additions of $N\times N$ matrices, and accordingly the computational complexity per iteration is $\ccalO(MN^3)$. The factor $M$ can be shaved off by parallelizing the computation of the gradient in Algorithm \ref{A:PGD}. Taking into account all these desirable features, Algorithm \ref{A:PGD} markedly improves upon its precursor in~\cite{RSSSAMGM_icassp17}.} Since multiple stationary points exist, we typically run Algorithm \ref{A:PGD} for $I$ random initializations.  Among the $I$ estimated filters we select the one whose eigenvectors lead to e.g., the sparsest graph shift $\bbS$ when solving \eqref{E:general_problem} with $f(\bbS)=\|\bbS\|_1$; see also the numerical tests in Section \ref{S:Simulations}. 

\begin{algorithm}[t]
	\caption{Graph filter identification using PGD}
	\label{A:PGD}
	\begin{algorithmic}[1]
		\STATE \textbf{Input:} $ \{ \bbC_{\bbx,m} , \hbC_{\bby,m}  \}_{m=1}^{M}$, step size $\eta > 0$, 
		tol. $\delta > 0$.
		\STATE \textbf{Initialize} $k=0$ and $\bbH_0 \in \ccalH_{N}$ at random.
		\REPEAT
		\STATE $\nabla \varepsilon(\bbH_k) \!=\! 4 \!\!\sum\limits_{m=1}^{M} \! \bbH_{k} \bbC_{\bbx,m} \bbH_{k}^T\bbH_{k} \bbC_{\bbx,m} - \hbC_{\bby,m} \bbH_{k} \bbC_{\bbx,m}$.
		\STATE $\bar{\bbH}_{k}= \left((\bbH_{k} - \eta \nabla\varepsilon(\bbH_k))+(\bbH_{k} - \eta \nabla\varepsilon(\bbH_k))^{T}\right)/2$.
		\STATE $\bbH_{k+1} = \bbH_{k}+\alpha_k(\bar{\bbH}_{k}-\bbH_{k})$, $\alpha_k$ chosen via line search.
		\STATE $k \gets k+1$.
		\UNTIL{$\lVert \bbH_{k}- \bbH_{k-1} \rVert_{F} \leq \delta$}
		\STATE \textbf{Return} $\hbH=\bbH_k$
	\end{algorithmic}
\end{algorithm}


\begin{remark}[Combining multiple sources of information] 
\normalfont The formulations in \eqref{E:linear_system} and \eqref{E:opt_filter_genericfilter_opt} can be combined if both input covariances and pairs of input-output realizations are available. It is also relevant in scenarios where the inputs are not zero mean but their first and second moments are known. Defining $\hbmu_{\bby,m}:=\frac{1}{P_m}\sum_{p=1}^{P_m} \bby_m^{(p)}$ and $\bbmu_{\bbx,m}:=\EE[\bbx_m]$, a natural cost would be $\tilde{\varepsilon}(\bbH)=\varepsilon(\bbH)+\nu\sum_{m=1}^M \|\hbmu_{\bby,m}-\bbH\bbmu_{\bbx,m}\|^2 $, where $\varepsilon(\bbH)$ is the cost function of \eqref{E:opt_filter_genericfilter_opt} and $\nu$ is a tuning parameter. PGD iterations silmilar to those in Algorithm \ref{A:PGD} can be derived to minimize the cost $\tilde{\varepsilon}(\bbH)$. 
\end{remark}

\subsection{Semidefinite relaxation}\label{Ss:SDR}

Here we show that the graph filter identification task can also be tackled using semidefinite relaxation (SDR)~\cite{luo2010semidefinite}, a convexification technique which has been successfully applied to a wide variety of non-convex quadratically-constrained quadratic programs (QCQP). 
To that end, we first cast the filter identification problem as a Boolean quadratic program (BQP){, see Appendix \ref{ProofAppBQP} for a proof}.
\begin{myproposition}\label{P:BQPt}
For $m=1,\ldots,M$ consider matrices $\bbA_{m} := (\bbC_{\bbx,m}^{-1/2}\bbV_{\mathbf{xyx},m}) \odot ( \bbC_{\bbx,m}^{-1/2}\bbC_{\mathbf{xyx},m}^{1/2}\bbV_{\mathbf{xyx},m}) \in \reals^{N^{2} \times N}$  and unknown binary vectors $\bbb_m\in\{-1,1\}^N$.  Define
$\bbPsi\in \reals^{N^{2}(M-1) \times NM}$ as 
\begin{equation} \label{e:matrix_A}
\bbPsi := \left[\begin{array}{cccccc}
\bbA_{1}&-\bbA_{2}&\mathbf{0}&\cdots &\mathbf{0}&\mathbf{0}\\
\mathbf{0}&\bbA_{2}&-\bbA_{3}&\cdots &\mathbf{0}&\mathbf{0}\\
\vdots&\vdots&\vdots&\ddots &\vdots&\vdots\\
\mathbf{0}&\mathbf{0}&\mathbf{0}&\cdots &\bbA_{M-1}&-\bbA_{M}\\ 
\end{array}\right] 
\end{equation}
and $\bbb:=[\bbb_1^T,\ldots,\bbb_{M}^T]^T\in \{-1,1\}^{NM}$.
If ${\normalfont{\text{rank}}}(\bbPsi )=NM-1$, then the filter can be exactly recovered (up to a sign) as ${\normalfont\textrm{vec}(\bbH^*)}= \bbA_1\bbb^{*}_{1}$,
where $\bbb^{*}_{1}$ is the first $N\times 1$ sub-vector of the solution to the following BQP problem
\begin{equation}
\bbb^* = \argmin_{\bbb \in \{-1,1\}^{NM}}
 \bbb^T\bbPsi^T\bbPsi \bbb. \label{e:BQP}
\end{equation}
\end{myproposition}

Problem \eqref{e:BQP} offers a natural formulation for the pragmatic setting whereby $\{\bbC_{\bby, m}\}_{m=1}^M$ are replaced by sample estimates, and one would aim at minimizing the residuals $\sum_{m=1}^{M-1} \| \hbA_m \bbb_m - \hbA_{m+1} \bbb_{m+1} \|^2=\|\hbPsi \bbb\|^2$ in the LS sense.  Given a solution of \eqref{e:BQP} with $\hbPsi$ replacing $\bbPsi$, $\bbH\in\ccalH_{N}$ can be estimated as [cf.~\eqref{E:filter_solutions_symmetric}]
\begin{equation} \label{e:estimate_filter_SDR}
\hbH= \frac{1}{M} \sum_{m=1}^{M} \bbC_{\bbx,m}^{-1/2}\hbC_{\mathbf{xyx},m}^{1/2}\hbV_{\mathbf{xyx},m}\diag(\bbb^{*}_{m})\hbV_{\mathbf{xyx},m}^T\bbC_{\bbx,m}^{-1/2}.
\end{equation}

Even though the BQP is a classical NP-hard combinatorial optimization problem~\cite{luo2010semidefinite}, via SDR one can obtain near-optimal solutions with provable approximation guarantees. To derive the SDR of \eqref{e:BQP}, first introduce the $NM\times NM$ symmetric PSD matrices $\bbW:=\bbPsi^T\bbPsi$ and $\bbB := \bbb \bbb^{T}$. By construction, the binary matrix $\bbB$ has rank one and its diagonal entries are $B_{ii} = b_{i}^{2} = 1$. Conversely, any matrix $\bbB\in\reals^{NM\times NM}$ that satisfies $\bbB \succeq 0$, $B_{ii}  = 1$, and $\text{rank}(\bbB) = 1$ necessarily has the form $\bbB = \bbb \bbb^{T}$, for some $\bbb \in \{-1,1\}^{NM}$. Using these definitions, one can write $\bbb^{T} \bbW \bbb = \text{trace}(\bbb^{T} \bbW \bbb) = \text{trace}(\bbW \bbb \bbb^{T}) = \text{trace}(\bbW \bbB)$ and accordingly \eqref{e:BQP} is equivalent to
\begin{align} \label{e:opt_prob_QP_equi}
\min_{\bbB}{} &{}
\text{trace}(\bbW \bbB)\\
\text{s. to } &{} \bbB  \succeq \bb0,\nonumber\\
& \text{rank}(\bbB) = 1,\nonumber\\
& B_{ii}=1, \quad i=1,\ldots,NM. \nonumber
\end{align}
The only source of non-convexity in \eqref{e:opt_prob_QP_equi} is the rank constraint, and dropping it yields the convex SDR
\begin{align}\label{e:opt_prob_QP_SDR} 
\bbB^*= \argmin_{\bbB }{} &{}
\text{trace}(\bbW \bbB)\\
\text{s. to } &{} \bbB  \succeq \bb0,\nonumber\\
& B_{ii}=1, \quad i=1,\ldots,NM. \nonumber
\end{align}
which coincides with the bidual (dual of the dual)  problem of \eqref{e:BQP}. Problem \eqref{e:opt_prob_QP_SDR} is a semidefinite program (SDP) and can be solved using an off-the-shelf interior-point method~\cite{ye1998interior}. 

\begin{algorithm}[t]
	\caption{Graph filter identification using SDR}
	\label{A:SDR}
	\begin{algorithmic}[1]
		\STATE \textbf{Input:} $ \bbW=\bbPsi^T\bbPsi \in \ccalH_{NM}$ and $L \in \naturals$.
		\STATE Solve the SDP in \eqref{e:opt_prob_QP_SDR} to obtain $\bbB^*$.
		\FOR{$l=1,\ldots,L$}
		\STATE Draw $\bbz_{l} \sim \ccalN(\bb0,\bbB^*)$.
		\STATE Round $\hbb_{l} = \text{sgn}(\bbz_{l})$.
		\ENDFOR
		\STATE Determine $l^*= \argmin_{l=1,\ldots,L} \hbb_{l}^{T} \bbW \hbb_{l}$.
		\STATE \textbf{Return} $\hbb_{l^*}$
	\end{algorithmic}
\end{algorithm}

It is immediate that a rank-one optimal solution $\bbB^*=\bbb^*(\bbb^*)^T$ of \eqref{e:opt_prob_QP_SDR} solves the original BQP as well; however, in general $\text{rank}(\bbB^*)\neq 1$. To generate a feasible solution of \eqref{e:BQP} from $\bbB^*$, we adopt the so-termed Gaussian randomization procedure~\cite{nesterov1998semidefinite,luo2010semidefinite}. 
The overall method is tabulated under Algorithm \ref{A:SDR} and the quality of the rounded solutions is evaluated via computer simulations in Section \ref{S:Simulations}. The computational complexity is discussed under Remark \ref{R:SDR_vs_PGD}.

Interestingly, it possible to derive theoretical approximation guarantees for the feasible solutions generated via the SDR scheme in Algorithm \ref{A:SDR}. {Leveraging a result in~\cite{nesterov1998semidefinite}, a guarantee for the BQP \eqref{e:BQP} follows immediately.}
%
%

\begin{mycorollary} \label{C:approximation}
	Let $\bbb^*$ be the solution of \eqref{e:BQP} and $\hbb_{l^*}$ be the output of Algorithm \ref{A:SDR}. For $\gamma=\left(1-\frac{2}{\pi}\right)\lambda_{\max}NM>0$, then
	\begin{equation} \label{e:Nesterov}
	\gamma+\frac{2}{\pi}(\bbb^{*})^{T} \bbW \bbb^* \geq \E{(\hbb_{l^*})^T \bbW\hbb_{l^*}} \geq (\bbb^{*})^{T} \bbW \bbb^*.
	\end{equation}
\end{mycorollary}

\noindent Notice that although the bounds in \eqref{e:Nesterov} offer guarantees in terms of the expected objective value,  particular realizations $\hbb_{l^*}$ tend to fall within those bounds with high probability if $L$ is chosen sufficiently large.

All in all, the recipe to estimate the graph filter via the SDR approach entails the following steps. First we calculate $\{\hbA_m\}_{m=1}^{M}$ from $\{\hbC_{\bby,m}, \bbC_{\bbx , m}\}_{m=1}^{M}$ using the expression in the statement of Proposition \ref{P:BQPt}, and form $\hbPsi$ as in \eqref{e:matrix_A} to finally obtain $\hbW = \hbPsi^{T} \hbPsi$. Next, a feasible solution $\hbb_{l^*}$ to the BQP is obtained after running Algorithm \ref{A:SDR} with $\hbW$ and an appropriate choice of $L$ as inputs. Finally, $\hbH$ is estimated using \eqref{e:estimate_filter_SDR}. 

\begin{remark}[SDR versus PGD] \label{R:SDR_vs_PGD}\normalfont
Although SDR has well-documented merits when dealing with non-convex BQPs (and other QCQPs) {in applications such as MIMO detection \cite{tan2001application} and transmit beamforming \cite{gershman2010convex}, it has so far not been explored for network topology inference.} The relaxation entails dropping a rank constraint after ``lifting'' a (binary) vector-valued problem with $NM$ variables to a matrix-valued one with $NM(NM+1)/2$ variables. This incurs an increase in memory and computational cost, since the complexity of a general-purpose interior point method to solve the resulting SDP is $\ccalO (N^7 M^7 \text{log}(1/\epsilon))$, for a prescribed solution accuracy $\epsilon > 0$~\cite{luo2010semidefinite}. The additional cost could hinder applicability of the SDR approach in Algorithm \ref{A:SDR} to problems involving very large networks. In those scenarios, the PGD iterations in Algorithm \ref{A:PGD} can still find stationary solutions with lower memory requirements and $\ccalO (M N^3)$ complexity per iteration. While nothing can be said a priori on the quality of the aforementioned stationary points, the more costly SDR-based solutions offer quantifiable approximation guarantees as asserted in Corollary \ref{C:approximation}. {Even though the focus here is not on pushing algorithmic scalability to the limit, the recovery performance of the proposed methods could serve as benchmark for faster (possibly randomized) algorithms that can handle larger graphs. Since the desired solution of \eqref{e:opt_prob_QP_SDR} has rank one, we envision future algorithmic improvements using large-scale SDP solvers based on e.g., the Burer-Monteiro factorization~\cite{Burer2003}.}
\end{remark}



\section{Numerical tests}\label{S:Simulations}

{We study the recovery of synthetic and real-world graphs to assess the performance of the proposed network topology inference algorithms. To this end, we specialize the general problem \eqref{E:general_problem} to recover a sparse adjacency matrix (thus $f(\bbS) = \| \bbS \|_1$ and $\ccalS = \ccalS_{\mathrm{A}}$) that is close in the Frobenius-norm sense to being diagonalized by the estimated eigenbasis $\hat{\bbV}$. Accordingly, given $\hbV$ we solve the following convex optimization problem
\begin{equation}\label{E:SparseAdj_l1_obj_noisy_matrix}
	\bbS^* := \argmin_{\bbLambda, \bbS \in \ccalS_\mathrm{A}} \
	\|\bbS\|_1,   \quad
	\text{s. to }\:\|\bbS-\hbV\bbLambda\hbV^T\|_F \leq \epsilon.
\end{equation}
A comprehensive numerical evaluation is carried out whereby we: (i)~study the graph inference performance in controlled synthetic settings (Section~\ref{Ss:performance_assess}); (ii)~carry out comparisons
with some state-of-the-art algorithms (Section~\ref{Ss:performance_comp}); (iii)~use this framework to gain insights about urban mobility patterns in New York City from data of Uber pickups in 2015 (Section~\ref{Ss:Uber}); and (iv)~cluster companies using a graph obtained from time series of their daily opening and closing stock prices (Section~\ref{Ss:Stocks}).}

\subsection{Performance assessment}\label{Ss:performance_assess}

Throughout this section, we infer networks from the observation of diffusion processes that are synthetically generated via graph filtering as in \eqref{E:Filter_input_output_time}. Denoting by $\bbS = \bbA$ the (possibly weighted) adjacency matrix of the sought undirected graph $\ccalG$, we consider two types of filters $\bbH_1 = \sum_{l=0}^{2}h_l \bbS^l$ and $\bbH_2 = (\mathbf{I} + \alpha \bbS)^{-1}$, where the coefficients $\{h_l\}$ and $\alpha$ are drawn uniformly at random on $[0,1]$. 
As a measure of recovery error, we adopt $\|{\bbS}^*-\bbS\|_F/\|\bbS\|_F$ (averaged over independent realizations of the experiment), where ${\bbS}^*$ is the solution of \eqref{E:SparseAdj_l1_obj_noisy_matrix} and $\bbS$ denotes the ground-truth GSO. {To directly assess edge-support recovery, we also compute the F-measure defined as the harmonic mean of edge precision and recall (precision is the percentage of correct edges in ${\bbS}^*$, and recall is the fraction of edges in $\bbS$ that are successfully retrieved).}

\noindent\textbf{Inference from input-output realization pairs.} We illustrate the method proposed in Section~\ref{Ss:linear_equations} by recovering a geographical network of the United States (US). More specifically, we consider a graph with $N=49$ nodes given by the 48 states in continental US plus the District of Columbia, and with unweighted edges indicating pairs of states that share a border.
We then generate $M$ i.i.d. random input-output pairs $\{\bbx_m,\bby_m\}_{m=1}^M$ as explained in Section~\ref{Ss:linear_equations}, where input signals (with uniformly-distributed entries in $[0,100]$) are diffused on the US graph via either the FIR filter $\bbH_1$ or the IIR filter $\bbH_2$. Given such data, the respective graph filters are estimated by forming \eqref{E:opt_filter_inputoutput_closedform}. 
Problem \eqref{E:SparseAdj_l1_obj_noisy_matrix} with $\hat{\bbV}$ given by the 
eigenvectors of the estimated filter is then solved in order to infer the graph.

\begin{figure}[t]
	\centering    
	{\includegraphics[width=0.95\linewidth]{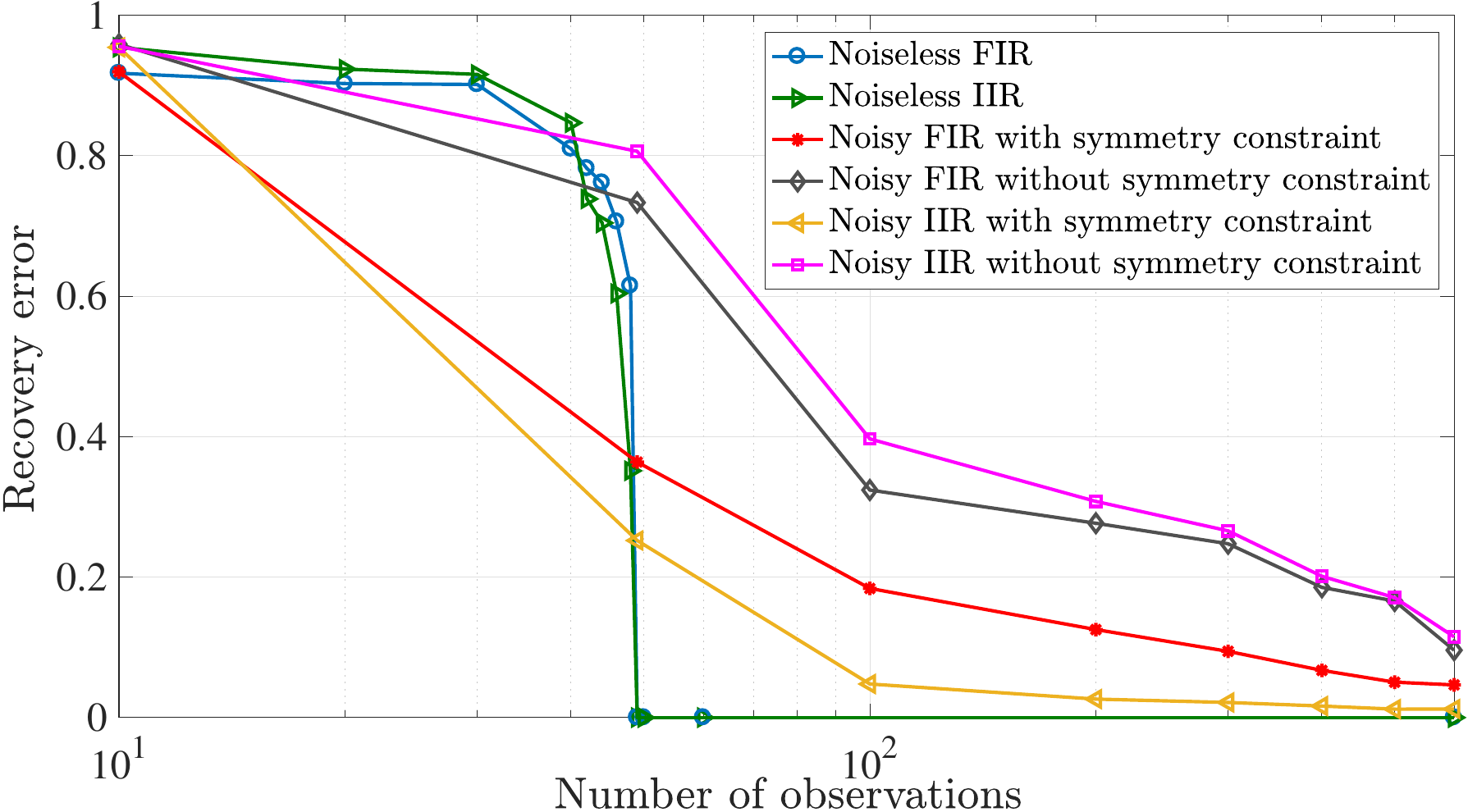}}\newline\\
	{\includegraphics[width=0.95\linewidth]{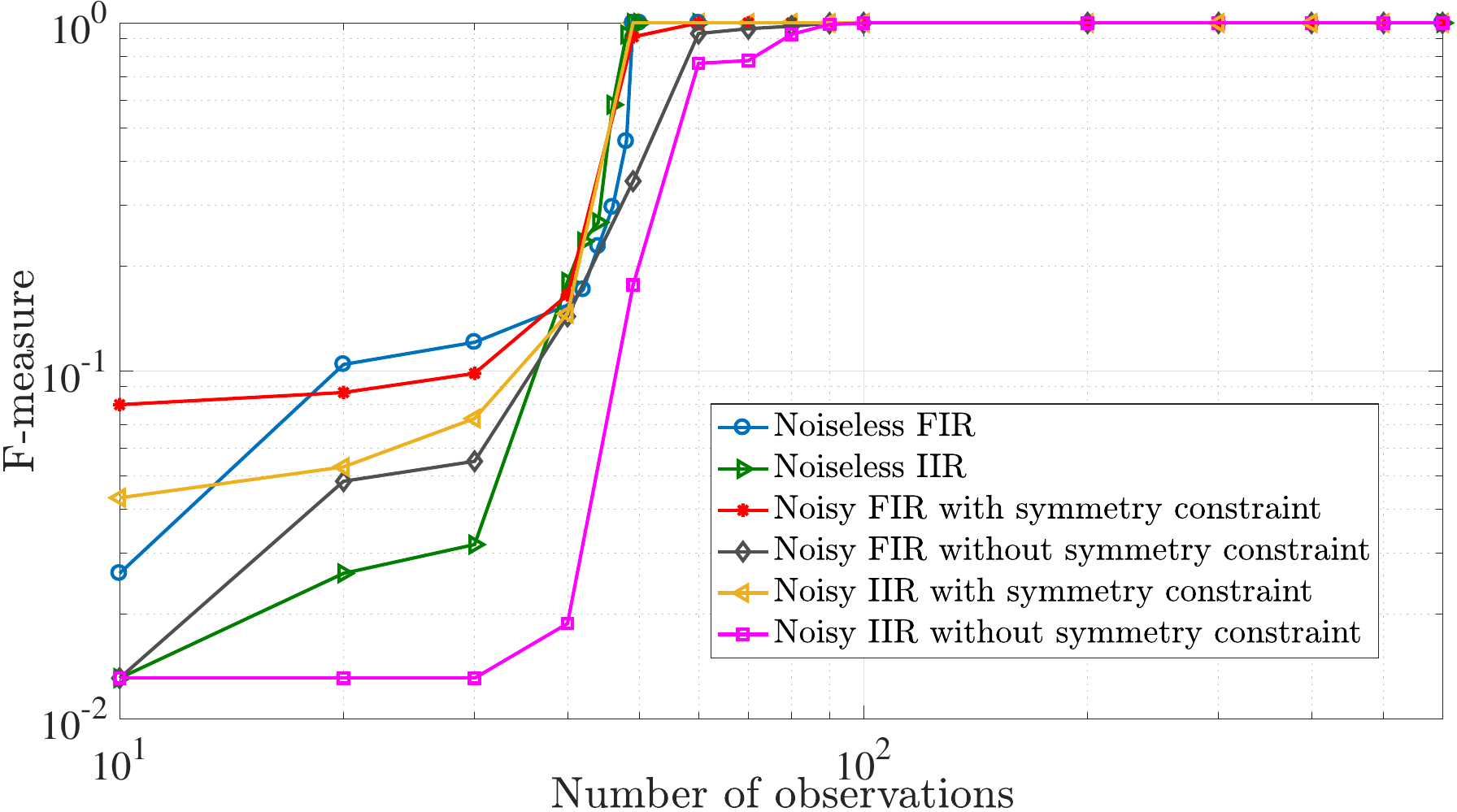}}
	\caption{Graph topology inference from input-output realization pairs. (top) Error in recovering a geographical graph of the US as a function of the number $M$ of noiseless and noisy input-output pairs observed for two different types of graph filters. When $M\geq N=49$ in the absence of noise, the graph is perfectly recovered. We discriminate two cases: one in which the symmetric structure of the filters is leveraged for recovery and one in which is not. {(bottom) Counterpart of the top plot but for the F-measure as figure of merit. Similar trends show the algorithm's success in recovering the GSO support.}}
	\label{fig:TSP_input_output}
	\vspace*{-0.5cm}
\end{figure}

In Fig.~\ref{fig:TSP_input_output} (top) we plot the recovery error as a function of $M$ for both types of filters.
First, notice that the performance is roughly independent of the filter type. 
More importantly, for $M \geq N$ in the absence of noise, the optimal filter minimizing \eqref{E:opt_filter_inputoutput_closedform} is unique (cf. Proposition~\ref{P:Closedform_inputoutputpairs}) and leads to perfect recovery. {Similar trends can be observed for the F-measure in Fig.~\ref{fig:TSP_input_output} (bottom), corroborating the efficacy of the approach in identifying the GSO support.
We also consider the case where observations of the output signals $\bby_m$ are corrupted by additive Gaussian noise with $-10$dB power. For this latter case, in Fig.~\ref{fig:TSP_input_output}  we also plot the error and F-measure when the symmetry of the filter is ignored and, thus, not leveraged to improve recovery performance (cf. discussion prior to Proposition~\ref{P:Closedform_inputoutputpairs}). First we notice that even though the estimation improves with increasing $M$, a larger number of observations is needed to guarantee successful recovery of the graph. Moreover, exploiting symmetry reduces the degrees of freedom in the filter to be inferred, thus markedly improving the recovery performance. Even though not shown in Fig. \ref{fig:TSP_input_output}, the performance was observed to degrade gracefully when the noise level increases.}

\noindent\textbf{Inference of PSD graph shifts.} We consider the karate club social network studied by Zachary~\cite{Zachary1977}, which is represented by a graph $\ccalG$ consisting of $N=34$ nodes or members of the club and undirected edges symbolizing friendships among them. Denoting by $\bbL_\mathrm{n}$ the normalized Laplacian of $\ccalG$, we define the graph-shift operator $\bbS \!=\! \bbI \!-\! \gamma \bbL_\mathrm{n}$ with $\gamma \!=\! 1/\lambda_{\max}(\bbL_\mathrm{n})$, modeling the diffusion of opinions between the members of the club. A graph signal $\bby$ can be regarded as a unidimensional opinion of each club member regarding a specific topic, and each application of $\bbS$ can be seen as an opinion update. Our goal is to recover $\bbL_\mathrm{n}$ -- hence, the social structure of the Karate club -- from the observations of opinion profiles. We consider $M$ different processes in the graph -- corresponding, e.g., to opinions on $M$ different topics -- and assume that an opinion profile $\bby_m$ is obtained by diffusing through the network an initial belief $\bbx_m$. More precisely, for each topic $m = 1, \ldots, M$, we model $\bbx_m$ as a zero-mean {Gaussian} process with known covariance $\bbC_{\bbx,m}$. {The input covariances are generated as $\bbC_{\bbx,m} = \mathbf{U}_{m} |\mathbf{\Lambda}_{m}| \mathbf{U}_{m}^{T}$, where the diagonal matrix $|\mathbf{\Lambda}_{m}|$ has diagonal entries equal to the absolute values of i.i.d. samples drawn from a standard normal distribution. Matrix $\mathbf{U}_{m} $ collects the eigenvectors of a symmetric matrix with i.i.d. standard normal entries.} We are then given a set $\{\bby^{(p)}_m\}_{p=1}^P$ of independent opinion profiles generated from different sources $\{\bbx^{(p)}_m\}_{p=1}^P$ diffused through filter $\bbH_1$ of unknown nonnegative coefficients. From these $P$ opinion profiles we first form an estimate $\hat{\bbC}_{\bby, m}$ of the output covariance. Leveraging that $\bbS$ is PSD and $h_l \geq 0$ for $l = 0,1,2$ (cf.~Section~\ref{Ss:PSD_filter}), then we estimate the filter $\bbH_1$ via \eqref{E:minimizer_PSD_closedform} and solve \eqref{E:SparseAdj_l1_obj_noisy_matrix} using the eigenbasis $\hat{\bbV}$ of the estimated filter. Set $\ccalS$ is modified accordingly for the recovery of a normalized Laplacian instead of an adjacency matrix; see \cite{segarra2016topoidTSP16}. 

\begin{figure}[t]
	\centering    
	{\includegraphics[width=0.85\linewidth]{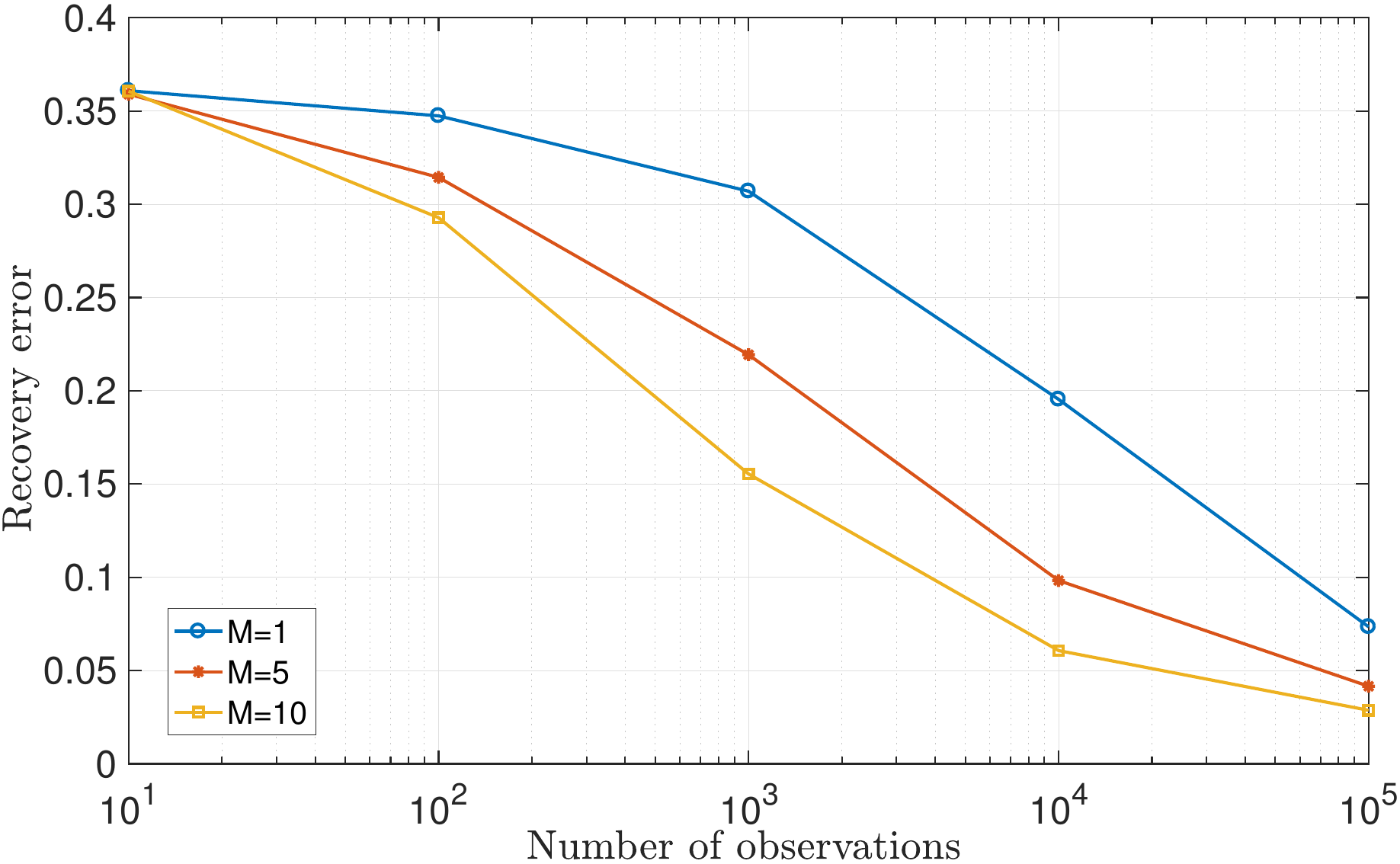}}
	\caption{Error in recovering Zachary's karate club social network as a function of the number $P$ of opinion profiles observed, and parametrized by the number of topics $M$. As expected, the graph estimate becomes increasingly accurate with increasing values of $P$ and $M$.}
	\label{fig:TSP_PSD}
	\vspace{-0.5cm}
\end{figure}

In Fig.~\ref{fig:TSP_PSD} we plot the recovery error as a function of the number of observations $P$ and for three different values of $M\in\{1,5,10\}$. As $P$ increases, the estimate $\hat{\bbC}_{\bby, m}$ becomes more reliable entailing a better estimation of the underlying filter and, ultimately, leading to more accurate eigenvectors $\hat{\bbV}$. Hence, we observe a decreasing error with increasing $P$. Moreover, for a fixed number of observations $P$, the error in the estimation of $\hat{\bbC}_{\bby, m}$ can be partially overcome by observing multiple diffusion processes, thus, larger values of $M$ lead to smaller graph recovery errors. {The results also confirm that, when only second-order statistical information is available, more observations are needed for reliable network inference relative to those required for the linear graph-filter identification task in the previous test case (cf. Fig. \ref{fig:TSP_input_output}).} 

{
\noindent\textbf{Inference of random graphs.} Here we evaluate the performance of Algorithm~\ref{A:PGD} on three different types of random graphs for varying number of nodes $N$. We consider the recovery of the adjacency matrix $\bbS = \bbA$ for: i) Erd\H{o}s-R\'enyi (ER) graphs with edge probability $p\!=\!0.3$; ii) Barab\'asi-Albert (BA) preferential attachment graphs generated from $m_{0}=0.5N$ initially placed nodes, where each new node is connected to $m=0.3N$ existing ones; and iii) small world (SW) graphs with rewiring probability $0.1$ and mean degree of $0.3N$. The parameter choices yield roughly equal mean degrees across all three graph models.

\begin{figure}[t]
	\centering    
	{\includegraphics[width=0.95\linewidth]{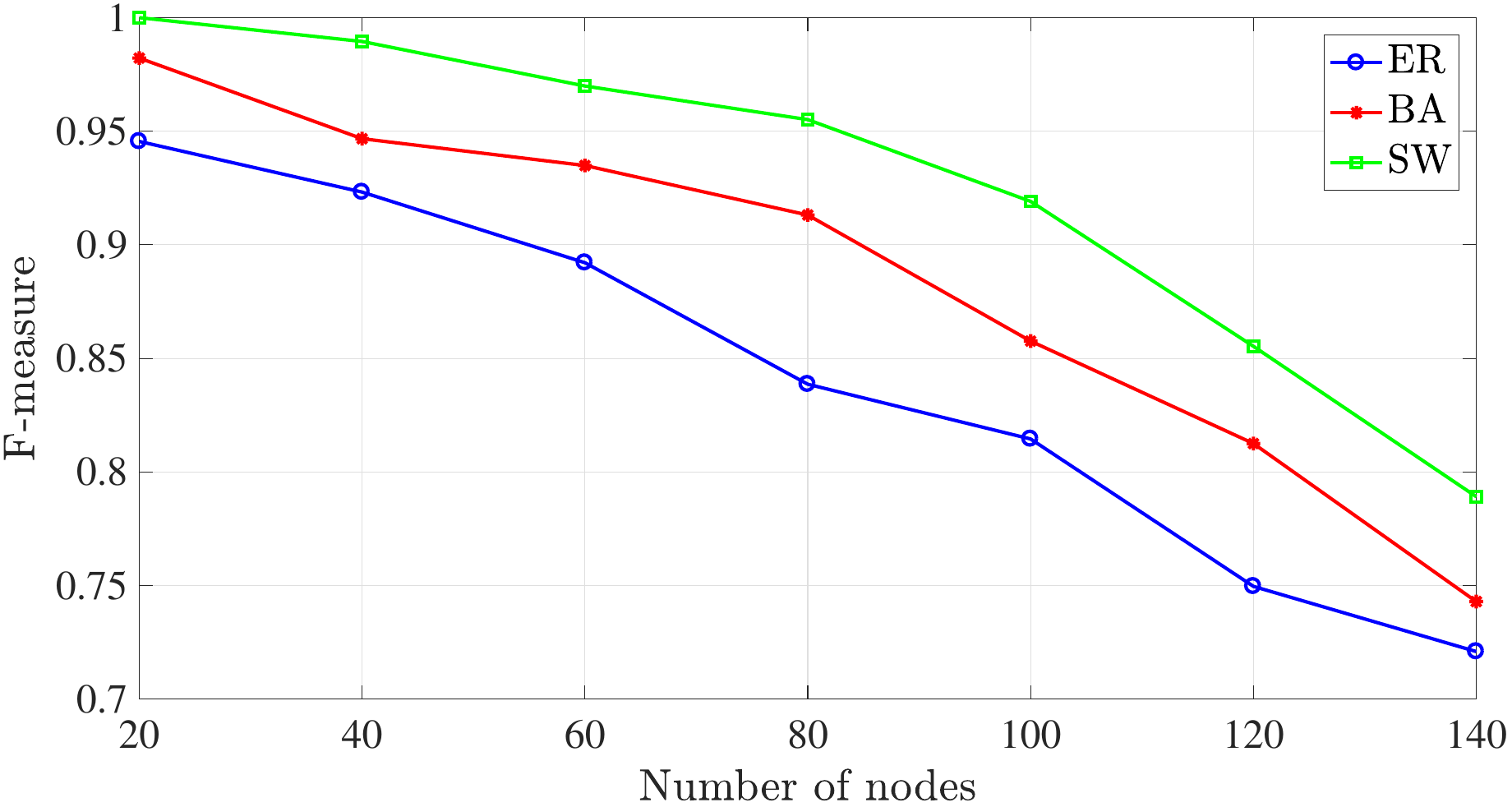}}
	\caption{{F-measure in recovering random graphs from $M=10$ number of processes and $P=10^6$ observations for each process. As number of nodes $N$ increases, the output covariances become less accurate and the performance of Algorithm~\ref{A:PGD} deteriorates. The SW graph with rewiring probability $0.1$, which is close to a regular graph, is the easiest to identify while the purely random ER graph is the hardest.} }
	\label{fig:random_graphs}
	\vspace*{-0.5cm}
\end{figure}

We consider $M=10$ different processes in the respective graphs, where we observe $10^{6}$ outputs for each process (i.e., $P = 10^{6}$). The given output signals $\{\bby^{(p)}_m\}_{p=1}^P$ are generated using the same procedure described under \emph{Inference of PSD graph shifts}. 
Then we estimate the filter $\bbH_1$ via Algorithm~\ref{A:PGD} and solve \eqref{E:SparseAdj_l1_obj_noisy_matrix} using the eigenbasis $\hat{\bbV}$ of the estimated filter. Fig.~\ref{fig:random_graphs} depicts the F-measure versus the number of nodes for the ER, BA, and SW random graphs. As $N$ increases for a fixed number of observations, the estimate $\hat{\bbC}_{\bby, m}$ becomes less reliable for larger graphs and as a result, the performance deteriorates. However, for $N\!\sim\!20-80$, we observe a reasonable recovery of the sought graphs. These observations are valid across all graph models. We also find that the SW graph with rewiring probability $0.1$, which is close to a regular graph, is the easiest to identify while the purely random ER graph is the hardest. For such unweighted graphs, this is aligned with findings in~\cite{segarra2016topoidTSP16}. The preferential attachment BA graph which relies on a copying procedure that introduces correlation~\cite[Ch. 6.4.1]{kolaczyk2009book}, falls in between the SW and ER ends of the spectrum. }

\noindent\textbf{Inference with a fixed signal budget.} In practice we estimate output covariances from observed signals via sample averaging. In particular, assume that we estimate the covariance of each of the $M$ processes by observing $P$ independent graph signals. 
For the cases where the total budget of signals $M\times P$ is fixed, this numerical test studies the trade-off between $M$ and $P$ as it pertains to recovery performance. Is it better to have accurate estimates of a few processes' covariances (larger $P$ and smaller $M$), or instead, coarser estimates of more processes (smaller $P$ and larger $M$)?
In order to answer this question, we {run an experiment whose goal is to recover the adjacency matrix of} the collaboration network of $N=31$ scientists working in the field of Network Science~\cite{newman2001scientific}, and model the diffusion of ideas among the scientists via simple linear dynamics as per $\bbH_1$. {The input signals are i.i.d., generated using the same procedure described under \emph{Inference of PSD graph shifts}.}  We observe outputs $\{\bby^{(p)}_m\}_{p=1}^P$ and implement Algorithm~\ref{A:SDR} for $L=10$ random draws to recover the collaboration graph from $M$ processes, each inferred from $P$ signals such that $M\times P$ is fixed. 

\begin{figure}[t]
	\centering    
	{\includegraphics[width=0.95\linewidth]{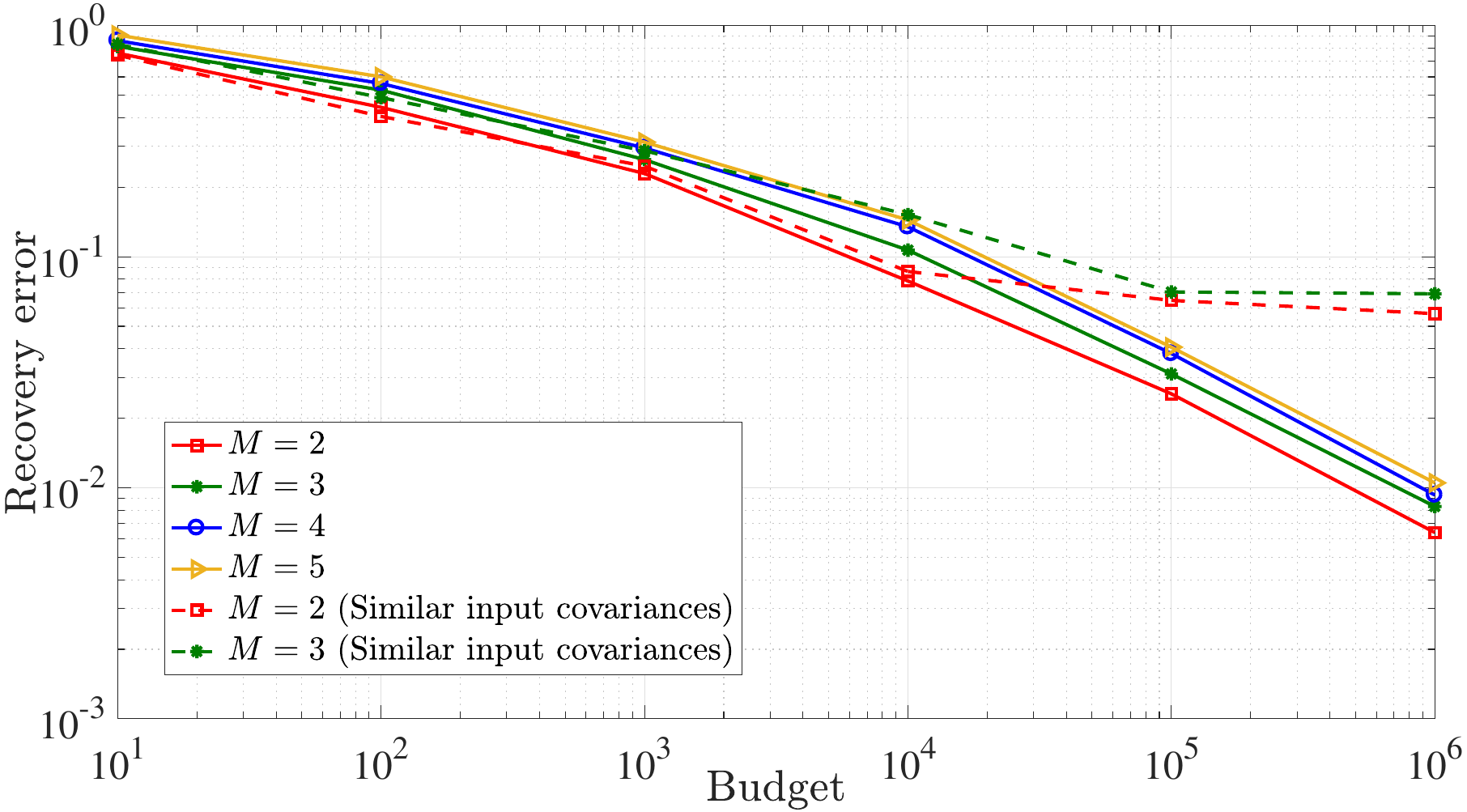}}
	\caption{{Error in recovering a collaboration network using SDR for varying budgets (number of processes $M$ times the signals observed per process $P$). For a fixed budget it is better to allocate the sensing resources in learning $M\!=\!2$ processes as accurately as possible, rather than having a coarser estimate of more ($M>2$) processes.}}
	\label{fig:limited_budget}
	\vspace*{-0.5cm}
\end{figure}

Fig.~\ref{fig:limited_budget} depicts the recovery error as a function of the total budget ($M\times P$) of observed signals, parametrized by $M\in\{2,3,4,5\}$. As expected, the error decreases for increasing budget for all values of $M$, since we can better estimate the covariances of all the processes. More interestingly, for a fixed budget it is better to consider only $M=2$ processes. Given that SDR can often recover the filter with perfect knowledge of $M=2$ covariances, it is better to focus on just two process and obtain the most accurate estimates of the associated covariances.

{Moreover, to emphasize the importance of the richness of the input processes, we repeat the experiment for a setting where the input covariances are similar to each other. 
To generate these covariances, we replace $\{\bbC_{\bbx,m}\}_{m=2}^{M}$ by $\bbC_{\bbx,1} + (10^{-8}) \{\bbC_{\bbx,m}\}_{m=2}^{M}$ for $M=2,3$; see the dashed lines in Fig.~\ref{fig:limited_budget} for the associated recovery performance.
Notice that for $M \leq 10^4$ the recovery error is comparable to (though slightly larger than) the corresponding counterparts in the original setting. 
More interestingly, for larger $M$ the recovery performance for the setting of similar covariance matrices saturates.
Intuitively, we know that for $M=1$ the problem is non-identifiable (cf. Section~\ref{Ss:symmetric_filter}). For practical purposes, the setting with similar covariance matrices behaves as having only one input process, thus resulting in an insurmountable error even for increasing values of $M$.}
%

{
\subsection{Performance comparison}\label{Ss:performance_comp}
Here we compare the performance of the proposed approach with common statistical approaches as well as with relevant counterparts in the GSP literature.

\noindent\textbf{Comparison with statistical approaches.} We analyze the performance of Algorithm~\ref{A:SDR} in comparison with two workhorse statistical methods, namely, (thresholded) correlation \cite[Ch. 7.3.1]{kolaczyk2009book} and graphical lasso \cite{GLasso2008}. Our goal is to recover the adjacency matrix of an undirected and unweighted graph with no self-loops. To that end, we are given (a varying number of) observed graph signals, which are modeled as the output of $M=2$  different diffusion processes. We test the recovery of adjacency matrices $\bbS=\bbA$ of ER graphs with $N = 20$ nodes and edge probability $p = 0.2$.
The zero-mean Gaussian inputs have covariance matrices $\bbC_{\bbx,m} = \mathbf{U}_{m} |\mathbf{\Lambda}_{m}| \mathbf{U}_{m}^{T}$, where the diagonal matrix $|\mathbf{\Lambda}_{m}|$ is generated as in \emph{Inference of PSD graph shifts}.
We consider two types of filters and accordingly $\bbU_m$: (i) the second-order filter $\bbH_1$ already defined. In this case, matrix $\mathbf{U}_{m} $ collects the eigenvectors of a symmetric matrix with i.i.d. standard normal entries; and (ii) filter $\bbH_3 = (\kappa \bbI + \bbS)^{-1/2} \bbC_{\bbx}^{-1/2}$, where $\kappa$ is selected to ensure that $\kappa \bbI + \bbS$ is positive definite and $\bbC_{\bbx}$ is the average of $\bbC_{\bbx,1}$ and $\bbC_{\bbx,2}$. In (ii), $\bbU_1=\bbU_2=\bbV $ are the eigenvectors of $\bbS$. This is to ensure that $\bbH_3$ is a polynomial on $\bbS$, i.e., a graph filter.

\begin{figure}[t] 
		\centering
		\includegraphics[width=0.95\linewidth]{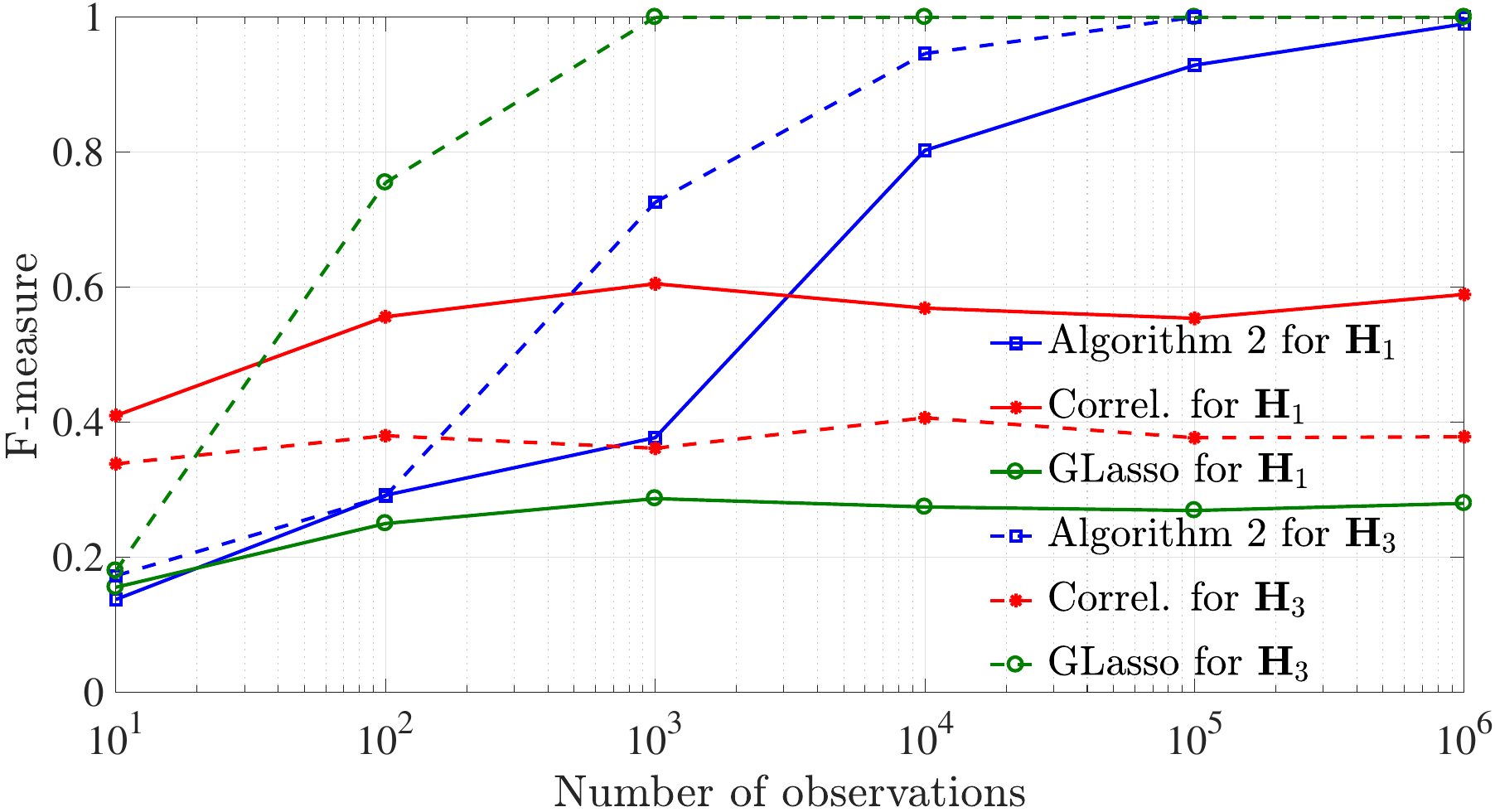}
	\caption{{Performance comparison in recovering random graphs between the proposed Algorithm~\ref{A:SDR}, graphical lasso, and correlation-based methods for general second-order filters ($\bbH_1$) and specific filters ($\bbH_3$) as a function of the number of observations for a fixed $M=2$.}}
	\label{fig:comparison_stat}
\end{figure}

We vary the number of diffused observed signals from $10$ to $10^{6}$ in powers of $10$. 
Output signals are generated by passing the inputs through a graph filter. 
For each combination of filter type and number of observed signals, we generate $10$ ER graphs that are used for training and $20$ ER graphs that are used for testing. 
Based on the $10$ training graphs, the optimal threshold for the correlation method and the regularization parameter for graphical lasso are determined and then used for the recovery of the $20$ testing graphs. 
In Fig.~\ref{fig:comparison_stat} we plot the performance of the three methods as a function of the number of filtered graph signals observed for filters $\bbH_1$ and $\bbH_{3}$, where each point is the mean F-measure over the $20$ testing graphs. When considering a general second-order graph filter $\bbH_1$, our proposed algorithm outperforms the other two baseline statistical methods. This is expected since the graph recovered by graphical lasso corresponds to the maximum-likelihood estimate of the precision matrix $\mathbf{C}_{\mathbf{y}}^{-1}=\bbH_1^{-1}\mathbf{C}_{\mathbf{x}}^{-1}\bbH_1^{-1}$, that bears no direct relation with $\bbS$.
However, for the specific case of graph filters $\bbH_3$, where the sought graph-shift operator matches the precision matrix in the off-diagonal entries, graphical lasso outperforms the other two which is due to the special design of graphical lasso to recover sparse precision matrices.

\noindent\textbf{Comparison with GSP methods.}  We finally compare the recovery performance of the proposed algorithms with the SEM inference method based on parallel factor (PARAFAC) tensor decomposition in~\cite{shen2017tensors}, the algorithm in~\cite{segarra2016topoidTSP16} which assumes that the observed graph processes are stationary, and a variant of the latter using a whitening transformation. 
We study a symmetric brain graph with $N=66$ nodes or neural regions and edge weights given by the density of anatomical connections between regions~\cite{hagmann2008mapping}.
Graph filter-based diffusion models with impulsive sources over the brain connectivity network [cf. \eqref{E:Filter_input_output_time}] were adopted and validated to model the progression of brain atrophy~\cite{hu_disease_brain}.
To compare with \cite[Algorithm~1]{shen2017tensors}, we draw zero-mean Gaussian inputs with diagonal covariance matrices $\bbC_{\bbx,m} = |\mathbf{\Lambda}_{m}| $, where $|\mathbf{\Lambda}_{m}|$ is generated as in \emph{Inference of PSD graph shifts}. We then generate output signals from the diffusion filter $\bbH_4 = (\bbI - \bbA)^{-1}$ (corresponding to a symmetric SEM), thus, output covariances satisfy $\mathbf{C}_{\mathbf{y},m}=\bbH_4\mathbf{C}_{\mathbf{x},m}\bbH_4$ for each $m$. Assuming perfect knowledge of the second-order statistics $\{ \mathbf{C}_{\mathbf{x},m} , \mathbf{C}_{\mathbf{y},m} \}_{m=1}^{M}$, 
we first estimate the graph filter using either the PGD approach in Algorithm~\ref{A:PGD} or the SDR approach in Algorithm~\ref{A:SDR}, and then solve \eqref{E:SparseAdj_l1_obj_noisy_matrix} to recover the adjacency matrix of the structural brain network.

\begin{figure}[t]
	\centering
	\includegraphics[width=0.95\linewidth]{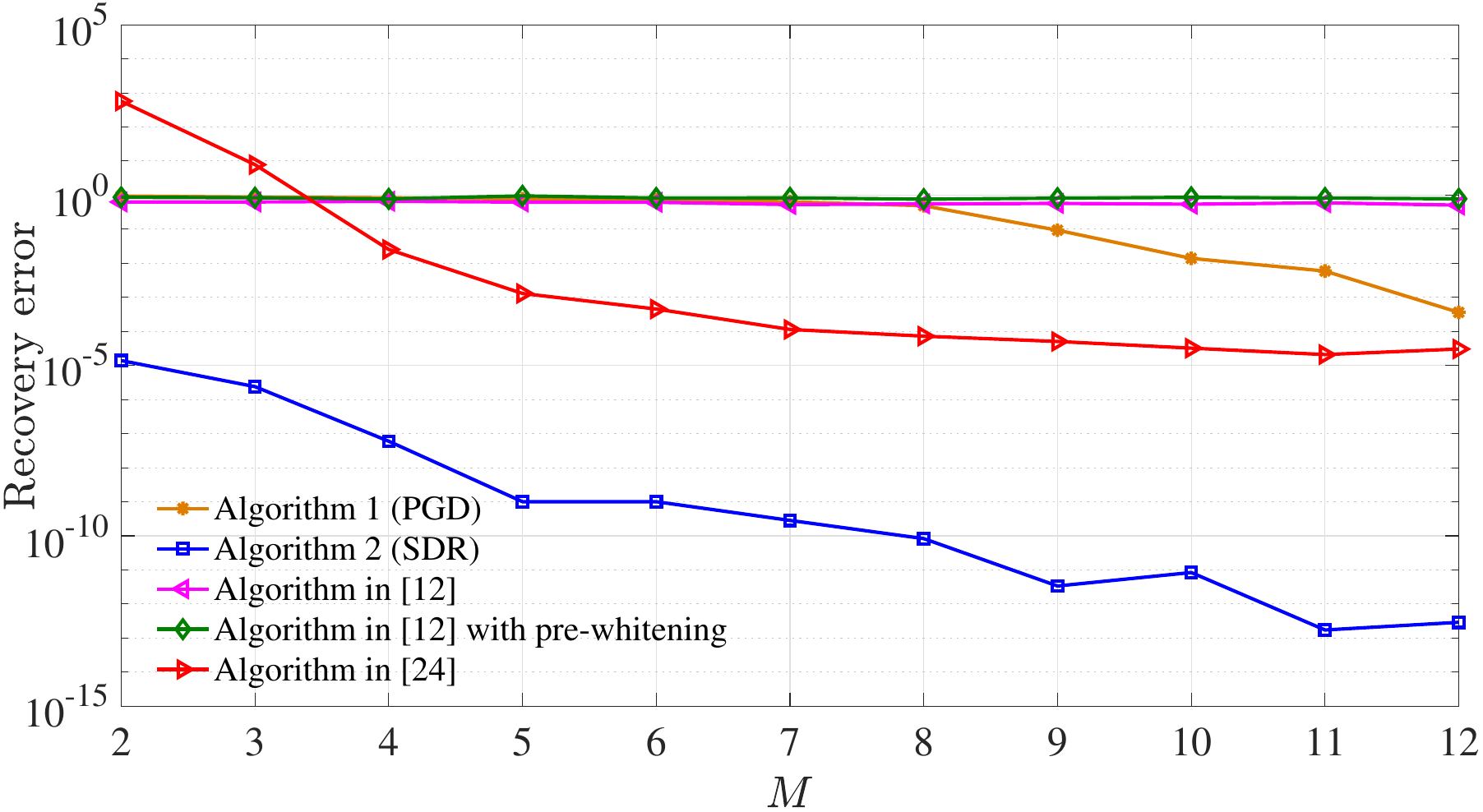}
	\caption{{Error in recovering a brain network as a function of the number $M$ of graph processes observed for different recovery algorithms. We assume perfect second-order statistical knowledge of the processes and a graph filter corresponding to a symmetric SEM. SDR outperforms PGD at the expense of a higher computational complexity (cf.~Remark~\ref{R:SDR_vs_PGD}). Also, \cite[Algorithm~1]{shen2017tensors} has a superior performance relative to the PGD algorithm due to the specific filter and set up. The mismatched stationarity assumption in~\cite{segarra2016topoidTSP16} explains its worse performance, even when using a whitening transformation as in \eqref{E:square_root_NS_one_sym}.}}
	 \label{fig:comparison_gsp}
	 \vspace*{-0.5cm}
\end{figure}

Fig.~\ref{fig:comparison_gsp} depicts the recovery error versus $M$ for: (i) Algorithms~\ref{A:PGD} and~\ref{A:SDR}; (ii) a baseline GSP-based method that exploits signal stationarity~\cite{segarra2016topoidTSP16}; (iii) a variant of the latter where we adopt a whitening transformation as in \eqref{E:square_root_NS_one_sym} and re-run the algorithm in \cite{segarra2016topoidTSP16} using the eigenvectors of $\bbC_{\bbx \bby \bbx}$ as spectral templates; and (iv) the tensor-based approach in~\cite{shen2017tensors}. 
First, we notice that the performance of both the PGD and SDR algorithms as well as \cite[Algorithm~1]{shen2017tensors} improves for increasing $M$. Moreover, the SDR approach uniformly (in $M$) outperforms all other methods. 
Recall that this gain in performance comes at the price of a higher computational complexity (relative to the PGD approach) as explained in Remark~\ref{R:SDR_vs_PGD} and is due to exploiting the structure of the solution in formulating the topology inference problem. Also, \cite[Algorithm~1]{shen2017tensors} outperforms the PGD algorithm, since it is tailored for SEMs matching this simulation setup. Finally, both proposed methods outperform the algorithm in~\cite{segarra2016topoidTSP16} (with or without pre-whitening) for all $M$. 
This is expected due to the model mismatch suffered when incorrectly assuming that the observed graph processes are stationary, and accordingly using the output covariance eigenvectors (or eigenvectors of $\bbC_{\bbx \bby \bbx}$, if whitened) as spectral templates of the recovered graph.
}

\subsection{Unveiling urban mobility patterns from Uber pickups}\label{Ss:Uber}

\begin{figure}[t]
	\centering    
	\includegraphics[width=0.9\linewidth]{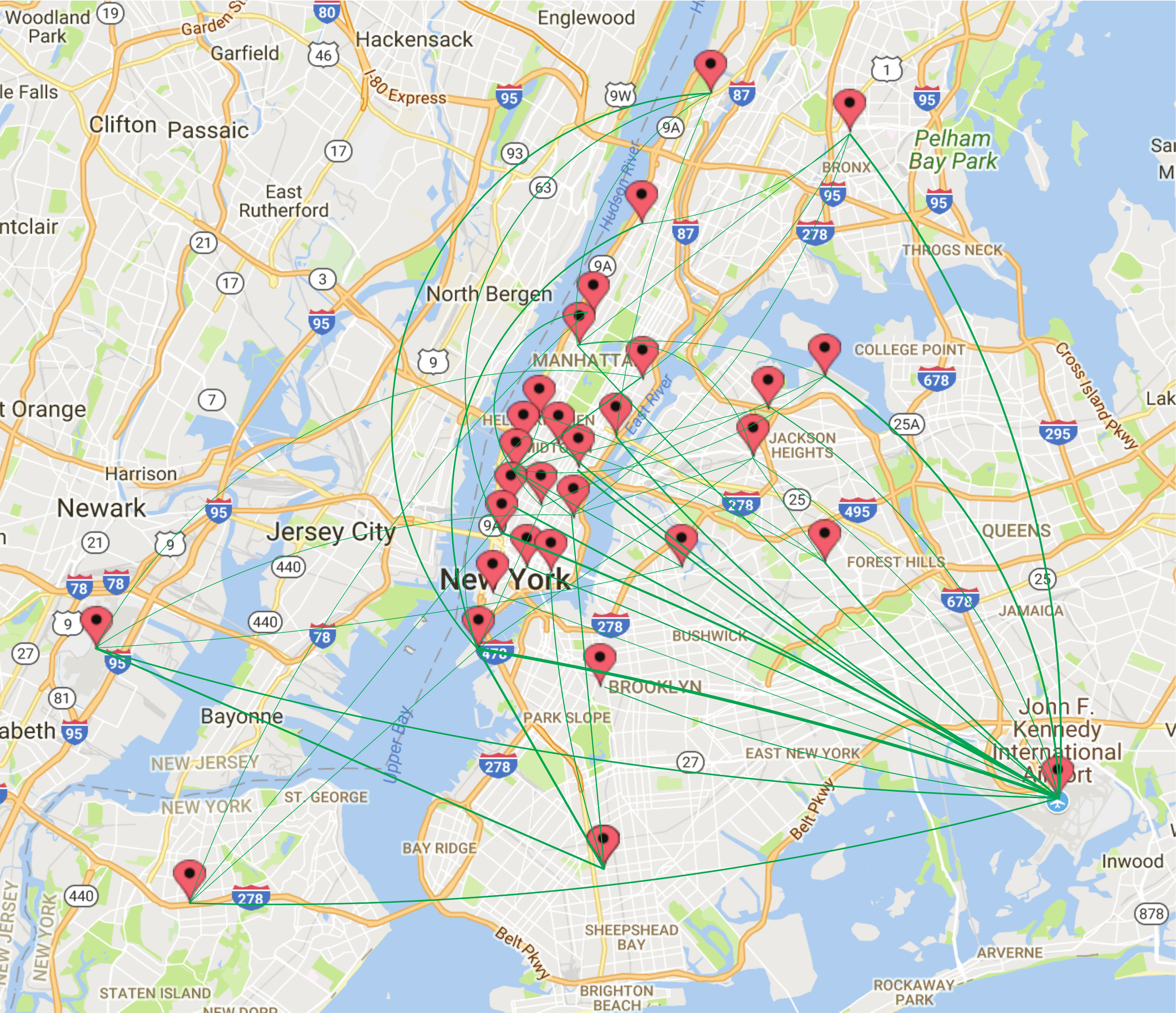}
	\caption{New York City's mobility pattern inferred from 2015 Uber pickups data. Most edges connect Manhattan with the other boroughs indicating that Uber is widely used to commute to/from the suburbs. Airports (Kennedy, Newark and LaGuardia) can also be distinguished as high degree nodes.}
	\label{fig:uber}
	\vspace*{-0.5cm}
\end{figure}

We implement our SDR graph topology inference method (Algorithm~\ref{A:SDR}) in order to detect mobility patterns in New York City from Uber pickups data\footnote{Dataset from \url{https://github.com/fivethirtyeight/uber-tlc-foil-response}}. We have access to times and locations of pickups from January 1\textsuperscript{st} to June 29\textsuperscript{th} 2015 for $263$ known location IDs. For simplicity, we cluster the locations into $N=30$ zones based on their geographical proximity; these are shown as red pins in Fig.~\ref{fig:uber}. These zones represent the nodes of the graph to be recovered. 
The total number of pickups aggregated by zone during a specific time horizon can be regarded as graph signals defined on the unknown graph. More specifically, we consider $M=2$ graph processes: weekday ($m=1$) and weekend ($m=2$) pickups. Moreover, we consider that the pickups from 6am to 11am constitute the inputs of our process whereas the pickups from 3pm to 8pm comprise the outputs of our process. To be more precise, for a specific day we aggregate all the pickups within 6-11am to form an input signal $\bbx$ and similarly we group all the pickups within 3-8pm to generate the associated output signal $\bby$. If the day considered is a weekday, we think of this pair as being generated from process $m=1$, and if it is a weekend we consider the pair coming from process $m=2$.  We repeat this procedure for all the days included in the period of study, and estimate input-output covariance pairs $\{\hat{\bbC}_{\bbx , m},\hat{\bbC}_{\bby, m}\}_{m=1}^{2}$.
We then run Algorithm~\ref{A:SDR} to infer an underlying graph filter $\hbH$ and solve~\eqref{E:SparseAdj_l1_obj_noisy_matrix} given the estimated eigenbasis of $\hbH$ to find a sparse mobility pattern. The modeling presumption is that throughout the day, the population diffuses over an unknown graph of mobility patterns we seek to identify. {By looking at aggregates over large number of trips, the population flows and mobility patterns in the city can be reasonably well approximated by linear diffusion dynamics; see also~\cite{thanou17}.}

The recovered graph is depicted in Fig.~\ref{fig:uber}, where the weights of the recovered edges is represented by the line widths in the figure. Given the nature of the input and output processes considered, the graph obtained is a sparse description of the mobility pattern of people throughout the day. Notice that most connections occur between Manhattan and the other boroughs (Queens, Bronx, Staten Island, Brooklyn and Newark), while only a few edges connect zones within Manhattan. This indicates that people use Uber to commute from their homes in the suburbs to their work (or leisure activities in the weekends) in the city. These findings are consistent with exploratory research of this same dataset~\cite{538blogpost} as well as a recent New York Times article that writes: \emph{``The ride-hail app has increasingly shifted its focus to the city's other four boroughs, where frustration over subway overcrowding and delays and fewer taxi options have made it the ride of choice for many. As a result, Uber is booming in the other boroughs, with half of all Uber rides now starting outside Manhattan\ldots''}~\cite{NYT2017}.
Lastly, observe that the JFK, Newark and LaGuardia airports are strongly connected with Manhattan and the other boroughs, as expected.

{
\subsection{Clustering firms from historical stock prices}\label{Ss:Stocks}

\begin{figure*}[t]
	\centering    
	\includegraphics[width=0.9\linewidth]{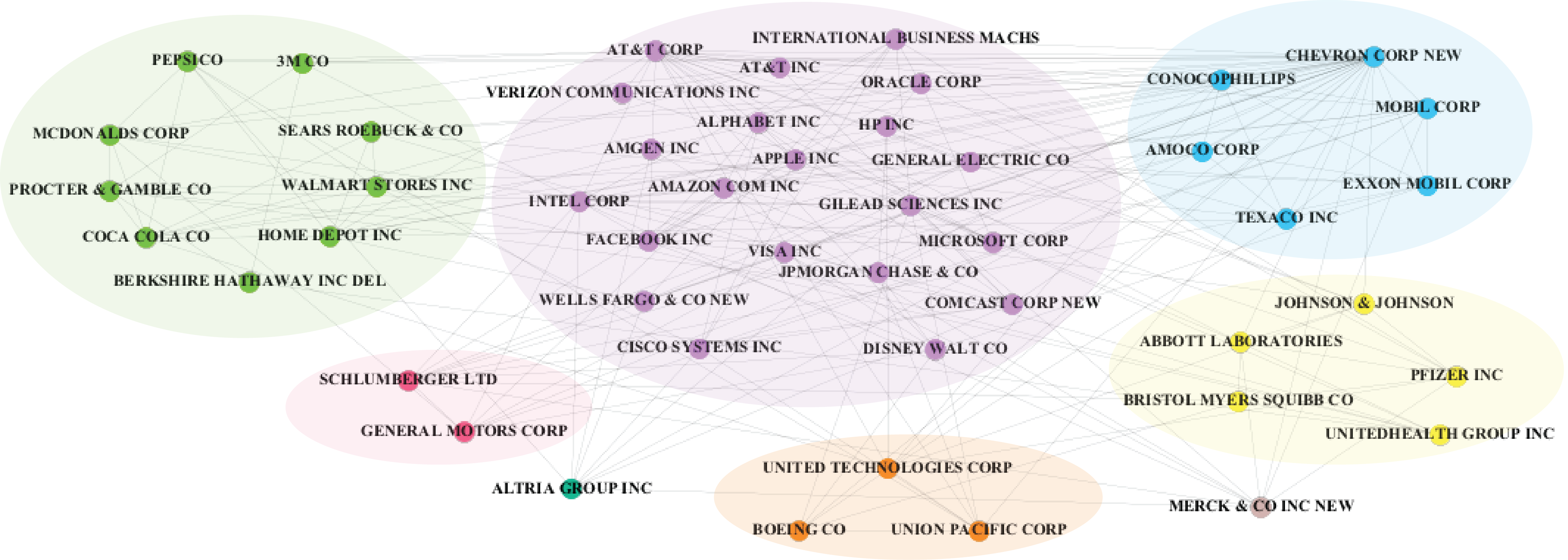}
	\caption{{The sparse network topology inferred from Yahoo's stock price data. Firms that represent similar sectors of the economy tend to be clustered together. For example, the blue community collects firms from the energy sector whereas the orange cluster relates to the aviation industry.}}
	\label{fig:stock}
	\vspace{-0.5cm}
\end{figure*} 

We implement our SDR topology inference approach on another real-world setting where we consider historical stock price data from Yahoo\footnote{Data from https://finance.yahoo.com/lookup?s=API}. 
Opening and closing prices are obtained from 18\textsuperscript{th} May 2012 to 30\textsuperscript{th} December 2016 for the top $50$ firms that have had the most wealth creation to shareholders in aggregate among all companies with common stock in the Center for Research in Security Prices (CRSP) database since July 1926 until December 2016 \cite[Table 5]{bessembinder2018stocks}\footnote{Of the $50$ firms mentioned, DuPont was excluded from the analysis for having excessive missing data and Pfizer and Warner Lambert were merged into a single one (since Pfizer bought Warner Lambert in 2000), resulting in a total of $N=48$ firms analyzed.}.
We consider the $N=48$ firms analyzed as nodes in a graph and the stock prices of the $1160$ days studied as graph signals defined on the graph. Our goal is then to recover the edges of this graph from the observed stock prices.
We consider $M=4$ graph processes for each three-month period of stock market activity. 
Moreover, the opening and closing prices of each day are respectively considered as inputs and outputs of the graph processes. 
A similar procedure to the one in Section~\ref{Ss:Uber} is used to estimate input-output covariance pairs $\{\hat{\bbC}_{\bbx , m},\hat{\bbC}_{\bby, m}\}_{m=1}^{4}$. 
We then run Algorithm~\ref{A:SDR} to infer an underlying graph filter $\hbH$ and solve~\eqref{E:SparseAdj_l1_obj_noisy_matrix} given the estimated eigenbasis of $\hbH$ to reveal a sparse graph of inter-dependencies among the $N$ companies.

The recovered graph is shown in Fig.~\ref{fig:stock}. 
In order to (indirectly) test the validity and usefulness of the obtained graph, we run the Louvain community detection algorithm~\cite{blondel2008fast} and analyze the resulting clusters, depicted in different colors in Fig.~\ref{fig:stock}. 
Notice that firms that broadly belong to the same economic sector are indeed clustered into the same communities.
For example, the blue community represents the firms that are mostly related to energy and oil industry, the purple community mostly relates to technology, telecommunication, and finance, the yellow community highly relates to healthcare and pharmaceutics, the orange community represents the aviation industry, and the green one embodies the food, drink, and retail sectors. }

\section{Concluding Summary}\label{S:Conclusions}

We studied the problem of inferring an undirected network from observations of \emph{non-stationary} signals diffused on the unknown graph. Relative to the stationary setting, the main challenge is that the GSO eigenvectors differ from those of the signal covariance matrix. To overcome this hurdle, we leverage that the sought eigenbasis is preserved by the polynomial graph filter that governs the diffusion process. As a result, the novel approach is to first identify the GSO eigenvectors from a judicious graph filter estimate, and then we rely on these spectral templates to estimate the eigenvalues by imposing desirable properties (such as edge sparsity) on the graph to be recovered. We propose different estimators of the symmetric diffusion filter depending on whether: i) explicit realizations; or, ii) second-order statistical information is available from the input-output graph signal pair. These estimators arise as solutions of systems of linear and quadratic matrix equations, respectively. We thus investigate identifiability properties of the resulting problems, and we develop PGD and SDR algorithms with complementary strengths to compute near-optimal solutions of the said system of quadratic equations.  The overall network topology inference pipeline is validated via comprehensive numerical tests on real social, author collaboration, and structural brain networks. {Moreover, we show how the developed graph inference tools can be utilized to unveil mobility patters in New York City from data of Uber pickups and to reveal inter-dependencies between firms from their historical stock prices.}

As future work, it is of interest to expand the scope of the proposed topology inference framework to accommodate directed graphs that represent the structure of signals (possibly) generated via nonlinear network interactions. Adaptive algorithms that can track slowly-varying graphs, and effect memory and computational savings by processing the signals on-the-fly are subject of ongoing investigation.

\appendix

\subsection{Proof of Proposition~\ref{P:Closedform_inputoutputpairs}}\label{ProofAppMinimizerInputOutput}
To show that a) is true, note first that the cost in \eqref{E:opt_filter_inputoutput_opt} can be compactly rewritten as $\|\bbY - \bbH\bbX\|_F^2$. Using the Kronecker product and the matrix vectorization operator, we can further rewrite it as $\|\bbY - \bbH\bbX\|_F^2=\|\text{vec}(\bbY)-\big(\bbX^T\otimes\bbI_{N}\big)\text{vec}(\bbH)\|_2^2$. Moreover, the redundant entries in $\text{vec}(\bbH)$ can be removed using the duplication matrix $\bbD_N$, to yield $\|\bbY - \bbH\bbX\|_F^2=\|\text{vec}(\bbY)-\big(\bbX^T\otimes\bbI_{N}\big)\bbD_N\text{vech}(\bbH)\|_2^2$.
This LS cost can be minimized using the Moore-Penrose pseudoinverse as
$\text{vech}(\bbH^*)=\left[\big(\bbX^T\otimes\bbI_{N}\big)\bbD_N\right]^\dagger\text{vec}(\bbY)$, so \eqref{E:opt_filter_inputoutput_closedform} follows. 

In order to prove that the b) holds true, we denote by $\{\bbv_i\}_{i=1}^{N-M_r}$ a basis of the null space $\text{ker}(\bbX^T)$. We use these vectors to form all non-repeated symmetric matrices of the form $\bbV_{ij} = \bbv_i \bbv_j^T + \bbv_j\bbv_i^T$, and then collect the $N_\bbH$ distinct entries of those symmetric matrices into vector $\tbv_{ij} = \bbD_N^\dagger \text{vec}(\bbV_{ij})=\bbD_N^\dagger(\bbv_i \otimes \bbv_j + \bbv_j \otimes \bbv_i)$.

\begin{mylemma}
Define the set $\ccalV^{\text{ker}}=\{\tbv_{ij}:\;\;i\leq j,\;\;1\leq j \leq N-M_r\}$, then it holds that: i) any of the $(N-M_r+1)(N-M_r)/2$ elements of $\ccalV^{\text{ker}}$ belongs to $\mathrm{ker}[(\bbX^T\otimes \bbI_N)\bbD_N]$; and ii) the elements in $\ccalV^{\text{ker}}$ are linearly independent. 
\end{mylemma}
\noindent \textbf{Proof:}
To establish i), we need to show that if $\tbv_{ij}\in \ccalV^{\text{ker}}$, then $[(\bbX^T\otimes \bbI_N)\bbD_N]\tbv_{ij}=\bbzero$. Indeed, we have that
\begin{align}
[(\bbX^T\otimes \bbI_N)\bbD_N]\tbv_{ij} ={} &\big((\bbX^T\otimes \bbI_N)\bbD_N \big) \bbD_N^\dagger \text{vec}(\bbV_{ij}) \nonumber\\
={}&(\bbX^T\otimes \bbI_N) \text{vec}(\bbV_{ij})\nonumber\\
 ={}&(\bbX^T\otimes \bbI_N) (\bbv_i \otimes \bbv_j + \bbv_j \otimes \bbv_i) \nonumber \\
={}&(\bbX^T\bbv_i\otimes \bbI_N\bbv_j) + (\bbX^T\bbv_j\otimes \bbI_N\bbv_i)\nonumber\\
={}&\bbzero,\nonumber
\end{align}
where the last equality follows because $\bbv_i, \bbv_j \in \mathrm{ker}(\bbX^T)$. 
	
To prove ii) we first define a matrix $\tbV^{\text{ker}}$ whose columns correspond to the vectors in $\ccalV^{\text{ker}}$ and write $\bbV^{\text{ker}} = \bbD_N \tbV^{\text{ker}}$. Notice that if $\bbV^{\text{ker}}$ is full column rank, then all the columns in $\tbV^{\text{ker}}$ must be linearly independent since $\bbD_N$ is just a replication operator. Hence, we need to show that vectors $\bbv_{ij}$ and $\bbv_{i'j'}$ (both columns of $\bbV^{\text{ker}}$) are orthogonal unless both $i=i'$ and $j=j'$. To see why this is true, note that $\bbv_{ij} = \bbv_i \otimes \bbv_j + \bbv_j \otimes \bbv_i$ and so
\begin{align}
	\bbv_{ij}^T\bbv_{i'j'}={} &\big[ (\bbv_i^T \! \otimes \! \bbv_j^T) + (\bbv_j^T \!\otimes \!\bbv_i^T)  \big]  \big[  (\bbv_{i'} \!\otimes \! \bbv_{j'}) + (\bbv_{j'} \!\otimes \!\bbv_{i'}) \big]\nonumber\\
	={} & (\bbv_i^T\bbv_{i'}\otimes \bbv_j^T\bbv_{j'}) + (\bbv_i^T\bbv_{j'}\otimes \bbv_j^T\bbv_{i'}) \nonumber\\
	&+ (\bbv_j^T\bbv_{i'}\otimes \bbv_j^T\bbv_{j'}) + (\bbv_j^T\bbv_{j'}\otimes \bbv_i^T\bbv_{i'})=\bbzero 
\nonumber	
\end{align}
since in all four summands there is at least one inner product that is zero, hence their sum vanishes as well.
 \hfill $\blacksquare$
 
\noindent From Lemma 1, we conclude that the dimension of $\mathrm{ker}(\bbX^T)$ is at least the cardinality of $\ccalV^{\text{ker}}$. Since $|\ccalV^{\text{ker}}|=(N-M_r+1)(N-M_r)/2$, it follows that $\mathrm{rank}[(\bbX^T\otimes \bbI_N)\bbD_N]$ is \textit{at most} $N_\bbH - (N-M_r+1)(N-M_r)/2$.

Finally, to see that c) is true, first notice that whenever $N < M_r$, then b) guarantees that $\rank(\big(\bbX^T\otimes\bbI_{N}\big)\bbD_N) < N_{\bbH}$. Consequently, the Moore-Penrose minimizer is not the unique LS minimizer. By contrast, if $N = M_r$ then $\bbX$ becomes full rank and, consequently, $(\bbX^T \otimes \bbI_N)$ is also full rank. Given that $\bbD_N$ is a full column rank matrix, $\big(\bbX^T\otimes\bbI_{N}\big)\bbD_N$ is also full column rank and the Moore-Penrose minimizer becomes the unique LS minimizer, as wanted.
\hfill $\blacksquare$

\subsection{Proof of Proposition~\ref{P:identifiability}}\label{ProofAppIdentifiabilityConvariances}

From~\eqref{E:quadratic_system} it follows that
\begin{equation*}
\bbC_{\bby,1}=\bbH \bbC_{\bbx,1} \bbH = \bbH \bbU \diag(\bblambda_1)\bbU^T \bbH = \bbQ \diag(\bblambda_1) \bbQ^T,
\end{equation*}
where we have implicitly defined $\bbQ := \bbH \bbU$. Notice that the basis $\bbU$ is completely determined since all eigenvalues in $\bblambda_1$ are distinct [cf.~A-1)]. Similarly, for the second diffusion process we obtain that $\bbC_{\bby,2}= \bbQ \diag(\bblambda_2) \bbQ^T$. Furthermore, if we define the matrix $\bbR_{\bbx} = [\bblambda_1,\bblambda_2]^T \in \reals^{2 \times N}$, and the $N \times N \times 2$ tensor $\underline{\bbC}_\bby$ (with slices along the third mode given by $\bbC_{\bby,1}$ and $\bbC_{\bby,2}$), then the partial symmetric PARAFAC decomposition of $\underline{\bbC}_\bby$ factors into the matrices $\bbQ$, $\bbQ$, and $\bbR_{\bbx}$; see \cite{shen2017tensors, parafac}. 

Recall that the \emph{Kruskal rank} of a matrix $\bbA \in \reals^{N \times M}$ (denoted by $\mathrm{kr}(\bbA)$) is defined as the maximum number $k$ such that \emph{any} combination of $k$ columns of $\bbA$ constitute a full rank submatrix. In this way, from condition A-2) it follows that $\mathrm{kr}(\bbR_{\bbx}) = 2$ and from the invertibility of $\bbH$ [cf.~A-4)] it follows that $\mathrm{kr}(\bbQ) = N$. Leveraging established results on the uniqueness of PARAFAC tensor decompositions (see \cite[Theorem 1]{shen2017tensors}), it follows that a PARAFAC decomposition of $\underline{\bbC}_\bby$ recovers $\bbQ$ and $\bbR_{\bbx}$ up to scaling and rotation ambiguities. However, given that we know $\bbR_{\bbx}$ a priori, part of those ambiguities can be resolved; see e.g. \cite[Lemma 1]{shen2017tensors}. 
To be more precise, it follows that we can recover $\bbQ'$, where $\bbQ' = \bbQ \diag(\bar{\bbb})$ for some unknown $\bar{\bbb} \in \{-1,1\}^N$. 
However, the following lemma establishes how to uniquely recover $\bar{\bbb}$, and uniqueness of $\bbH = \bbQ \bbU^T = \bbQ' \diag(\bar{\bbb}) \bbU^T$ follows.

\begin{mylemma}\label{L:appendix}
Vector $\bar{\bbb} \in \{-1,1\}^N$ can be found as the only vector (up to sign) such that $\bbQ' \diag(\bar{\bbb}) \bbU^T$ is symmetric.
\end{mylemma}
\noindent \textbf{Proof:}
Combining the facts that $\bbH = \bbQ' \diag(\bar{\bbb}) \bbU^T$ for the true $\bar{\bbb}$ and that $\bbH$ is symmetric, it follows that $\bbQ' \diag(\bar{\bbb}) \bbU^T$ is symmetric. Thus, we are left to show that no other $\bbb' \in \{-1,1\}^N$ leads to a symmetric $\bbQ' \diag(\bbb') \bbU^T$. To show this, begin by defining the symmetric matrix $\bbP := \bbU \diag(\bar{\bbb}) \diag(\bbb') \bbU^T$. Hence it follows that if indeed $\bbQ' \diag(\bbb') \bbU^T = (\bbQ' \diag(\bbb') \bbU^T)^T$ then
\begin{equation}
\bbH \bbU \diag(\bar{\bbb}) \diag(\bbb') \bbU^T = \bbU \diag(\bar{\bbb}) \diag(\bbb') \bbU^T \bbH.
\end{equation}
This means that $\bbH$ and $\bbP$ must commute and this requires them to be simultaneously diagonalizable. However, since $\bbU$ and $\bbV$ (the eigenbasis of $\bbH$) do not share any eigenvector [cf.~A-3)], this can only happen if $\bbP = \bbI$ or $\bbP = - \bbI$. Hence, we must have that $\bbb' = \bar{\bbb}$ or $\bbb' = - \bar{\bbb}$ and identifiability of $\bar{\bbb}$ is guaranteed.
 \hfill $\blacksquare$
 	

{
\subsection{Proof of Proposition~\ref{P:BQPt}}\label{ProofAppBQP}

Using properties of the Khatri-Rao product, feasible graph filters $\bbH\in\ccalH_{M}^{\text{sym}}$ satisfy the  system of equations [cf.~\eqref{E:solution_set}]
	\begin{equation} \label{E:A_m_b_m_H_hat}
		\bbA_m \bbb_m = \text{vec}(\bbH), \quad m=1,\ldots,M,
	\end{equation}
	with matrices $\bbA_{m}$ defined in the statement of the proposition and for some binary vectors
	$\bbb_m \in \{-1,1\}^N$. Based on \eqref{e:matrix_A}, the equations in \eqref{E:A_m_b_m_H_hat} can be
	compactly and equivalently rewritten as 
	\begin{equation} \label{E:A_b_0}
		\bbPsi \bbb= \bb0, \quad \bbA_1 \bbb_1=\text{vec}(\bbH).
	\end{equation}
	Under the assumption that the covariances $\{\bbC_{\bby, m}\}_{m=1}^N$ are perfectly known, then the filter $\bbH$ can be uniquely identified (up to a sign ambiguity) provided $\text{rank}(\bbPsi)=NM-1$. This way, $\mathrm{ker}(\bbPsi)$ has dimensionality one and the basis vector $\bbb^*=[(\bbb_1^*)^T,\ldots,(\bbb_{M}^*)^T]^T\in\{-1,1\}^{NM}$ of the aforementioned null space can be used to recover the filter via $\bbA_1 \bbb_1^*=\text{vec}(\bbH)$. Moreover, the desired solution of \eqref{E:A_b_0} can be equivalently obtained from the  BQP
	\begin{equation*}
		\bbb^* = \argmin_{\bbb \in \{-1,1\}^{NM}}
		\|\bbPsi \bbb\|^2,
	\end{equation*}
	which is identical to \eqref{e:BQP}. \hfill$\blacksquare$
}

{
\section*{Acknowledgment}

We would like to thank the authors of~\cite{shen2017tensors} for sharing the Matlab code of their algorithm, and the anonymous reviewers for constructive comments on an earlier version of this paper.} 

\bibliographystyle{IEEEtran}
%
\bibliography{citations}

\end{document}